\documentclass[fleqn,twoside]{article}
\usepackage[headings]{espcrc2_arXiv}
\usepackage{graphicx,natbib}
\newcommand{\gsim}{\lower.4ex\hbox{$\;\buildrel >\over{\scriptstyle\sim}\;$}}
\newcommand{\lsim}{\lower.4ex\hbox{$\;\buildrel <\over{\scriptstyle\sim}\;$}}
\newcommand{\arcsec}{\ensuremath{^{\prime\prime}}}
\newcommand{\arcmin}{\ensuremath{^{\prime}}}
\newcommand{\etal}{et al.}

\title{Solar System Science with SKA}

\author{
   B.J. Butler\address{NRAO, Socorro, NM, USA, bbutler@nrao.edu},
   D.B. Campbell\address{Cornell University, Ithaca, NY, USA},
   I. de Pater\address{University of California at Berkeley, Berkeley, CA, USA},
   D.E. Gary\address{New Jersey Institute of Technology, Newark, NJ, USA}
   }

\runtitle{Solar System Science with SKA}
\runauthor{B.J. Butler \etal}

\begin{document}

\begin{abstract}
Radio wavelength observations of solar system bodies reveal unique 
information about them, as they probe to regions inaccessible by 
nearly all other remote sensing techniques and wavelengths.  As such, 
the SKA will be an important telescope for planetary science studies.  
With its sensitivity, spatial resolution, and spectral flexibility and
resolution, it will be used extensively in planetary studies.  It will 
make significant advances possible in studies of the deep atmospheres, 
magnetospheres and rings of the giant planets, atmospheres, surfaces, 
and subsurfaces of the terrestrial planets, and properties of small 
bodies, including comets, asteroids, and KBOs.  Further, it will allow 
unique studies of the Sun.  Finally, it will allow for both indirect 
and direct observations of extrasolar giant planets.
\vspace*{1pc}
\end{abstract}

\maketitle

\section{Introduction}

Radio wavelength observations of solar system bodies are an important 
tool for planetary scientists.  Such observations can be used to 
probe regions of these bodies which are inaccessible to all other 
remote sensing techniques.  For solid surfaces, depths of up to meters 
into the subsurface are probed (the rough rule of thumb is that depths 
to $\sim$10 wavelengths are sampled).  For giant planet atmospheres, 
depths of up to 10's of bars are probed.  Probing these depths yields 
unique insights into the bodies, their composition, physical state, 
dynamics, and history.  The ability to resolve this emission is 
important in such studies.  The VLA has been the state-of-the-art 
instrument in this respect for the past 20 years, and its power is 
evidenced by the body of literature in planetary science utilizing 
its data.  With its upgrade (to the EVLA), it will remain in this 
position in the near future.  However, even with that upgrade, there 
are still things beyond its capabilities.  For these studies, the SKA 
is the only answer.  We investigate the capabilities of SKA for 
solar system studies below, including studies of the Sun.  We also 
include observations of extrasolar giant planets.  Such investigations 
have appeared before (for example, in the EVLA science cases in a 
general sense, and more specifically in \citet{depa99}), and we build 
on those previous expositions here.

\section{Instrumental Capabilities}

For solar system work the most interesting frequencies in most cases 
are the higher ones, since the sources are mostly blackbodies to first 
order (see discussion below for exceptions).  We are very interested 
in the emission at longer wavelengths, of course, but the resolution 
and source detectability are maximized at the higher frequencies.  
To frame the discussion below, we need to know what those resolutions 
and sensitivities are.  We take our information from the most recently 
released SKA specifications \citep{jone04}.

The current specifications for SKA give a maximum baseline of 3000 km.  
Given that maximum baseline length, Table~\ref{resntable} shows 
the resolution of SKA at three values of the maximum baseline, 
assuming we can taper to the appropriate length if desired.  In 
subsequent discussion we will translate these resolutions to physical 
dimensions at the distances of solar system bodies.

\begin{table}[tbh]
\caption{Resolution in masec for SKA.}
\label{resntable}
\vspace*{1truemm}
\footnotesize
\begin{center}
\begin{tabular}{cccc}
\hline\hline
\noalign{\vspace{3pt}}
$\nu$ (GHz) & $\theta_{300}$ & $\theta_{1000}$ & $\theta_{3000}$ \\
\noalign{\vspace{3pt}}
\hline
\noalign{\vspace{3pt}} 
0.5 & 410 & 120 & 40 \\
1.5 & 140 &  40 & 14 \\
5   &  40 &  12 &  4 \\
25  &   8 &   3 &  1 \\
\noalign{\vspace{3pt}}
\hline\hline
\end{tabular}
\end{center}
\end{table}

The specification calls for $A/T$ of 5000 at 200 MHz; 20000 from 500 
MHz to 5 GHz; 15000 at 15 GHz; and 10000 at 25 GHz.  The specification 
also calls for 75\% of the collecting area to be within 300 km.  Let us 
assume that 90\% of the collecting area is within 1000 km.  The 
bandwidth specification is 25\% of the center frequency, up to a 
maximum of 4 GHz, with two independently tunable passbands and in each 
polarization (i.e., 16 GHz total bandwidth at the highest frequencies). 
Given these numbers, we can then calculate the expected flux density 
and brightness temperature noise values, as shown in 
Table~\ref{noisetable}.

\begin{table*}[tbh]
\caption{Sensitivities for SKA in nJy and K in 1 hour of observing.}
\label{noisetable}
\vspace*{1truemm}
\footnotesize
\begin{center}
\begin{tabular}{ccccccc}
\hline\hline
\noalign{\vspace{3pt}}
$\nu$ (GHz) & ${\Delta}F_{300}$ & ${\Delta}T_{B_{300}}$ &
   ${\Delta}F_{1000}$ & ${\Delta}T_{B_{1000}}$ &
   ${\Delta}F_{3000}$ & ${\Delta}T_{B_{3000}}$ \\
\noalign{\vspace{3pt}}
\hline
\noalign{\vspace{3pt}} 
0.5 & 97 & 2.3 & 81 & 22  & 73 & 170 \\
1.5 & 56 & 1.3 & 47 & 12  & 42 & 100 \\
5   & 31 & 0.7 & 26 & 6.8 & 23 &  55 \\
25  & 34 & 0.8 & 29 & 7.6 & 26 &  62 \\
\noalign{\vspace{3pt}}
\hline\hline
\end{tabular}
\end{center}
\end{table*}

\section{Giant Planets}

Observations of the giant planets in the frequency range of SKA are 
sensitive to both thermal and nonthermal emissions.  These emissions 
are received simultaneously, and can be distinguished from each other 
by examination of their different spatial, polarization, time (e.g., 
for lightning), and spectral characteristics.  Given the sensitivity 
and resolution of SKA (see Table~\ref{giantresn}), detailed images of 
both of these types of emission will be possible.  We note, however, 
the difficulty in making images with a spatial dynamic range of $>$ 
1000 (take the case of Jupiter, with a diameter of 140000 km, and 
resolution of $\sim$100 km) - this will be challenging, not only in the 
measurements (good short spacing coverage - down to spacings of order 
meters - is required), but in the imaging itself.

\begin{table}[tbh]
\caption{SKA linear resolution for giant planets.}
\label{giantresn}
\vspace*{1truemm}
\footnotesize
\begin{center}
\begin{tabular}{cccc}
\hline\hline
\noalign{\vspace{3pt}}
     & Distance & \multicolumn{2}{c}{resolution (km) $^*$} \\
Body & (AU)     & $\nu=$2 GHz & $\nu=$20 GHz \\
\noalign{\vspace{3pt}}
\hline
\noalign{\vspace{3pt}} 
Jupiter & 5  & 120 & 10  \\
Saturn  & 9  & 210 & 20  \\
Uranus  & 19 & 420 & 40  \\
Neptune & 30 & 690 & 70  \\
\noalign{\vspace{3pt}}
\hline\hline
\end{tabular}
\end{center}
\ 

$^*$ assuming maximum baseline of 1000 km
\end{table}

\subsection{Nonthermal emission}

Nonthermal emissions from the giant planets at frequencies between 0.15 
and 20 GHz are limited to synchrotron radiation and atmospheric 
lightning. Both topics have been discussed before in connection to SKA 
by \citet{depa99}. We review and update these discussions here.

\subsubsection{Synchrotron radiation}

Synchrotron radiation results from energetic electrons ($\sim$ 1-100 
MeV) trapped in the magnetic fields of the giant planets.  At present, 
synchrotron emission has only been detected from Jupiter, where 
radiation at wavelengths longer than about 6 cm is dominated by this 
form of emission \citep{berg76}.  Saturn has no detectable synchrotron 
radiation because the extensive ring system, which is almost aligned 
with the magnetic equatorial plane, absorbs energetic particles 
\citep{mcdo80}.  Both Uranus and Neptune have relatively weak magnetic 
fields, with surface magnetic field strengths $\sim$20-30 times weaker 
than Jupiter.  Because the magnetic axes make large angles 
(50-60$^\circ$) with the rotational axes of the planets, the 
orientation of the field of Uranus with repect to the solar wind is in 
fact not too dissimilar from that of Earth (because its rotational 
pole is nearly in the ecliptic), while the magnetic axis of Neptune 
is pointed towards the Sun once each rotation period.  These profound
changes in magnetic field topology have large effects on the motion of 
the local plasma in the magnetosphere of Neptune.  It is unclear if 
there is a trapped population of high energy electrons in the radiation 
belt of either planet, a necessary condition for the presence of 
synchrotron radiation.  Before the Voyager encounter with the planet, 
\citet{depa89} postulated the presence of synchrotron radiation from 
Neptune.  Based on the calculations in their paper, the measured 
magnetic field strengths, and 20-cm VLA observations 
\citep[see, e.g.,][]{depa91b} we would estimate any synchrotron 
radiation from the two planets not to exceed $\sim$0.1 mJy.  This, or 
even a contribution one or two orders of magnitude smaller, is trivial 
to detect with the SKA.  It would be worthwhile for the SKA to search 
for potential synchrotron emissions off the disks of Uranus and Neptune 
(and SKA can easily distinguish the synchrotron emission from that from 
the disk based on the spatial separation), since this information would 
provide a wealth of information on the inner radiation belts of these 
planets.

\begin{figure*}[tbh]
\centering
\begin{minipage}[c]{.45\textwidth}
   \centering
   \includegraphics[width=0.9\textwidth]{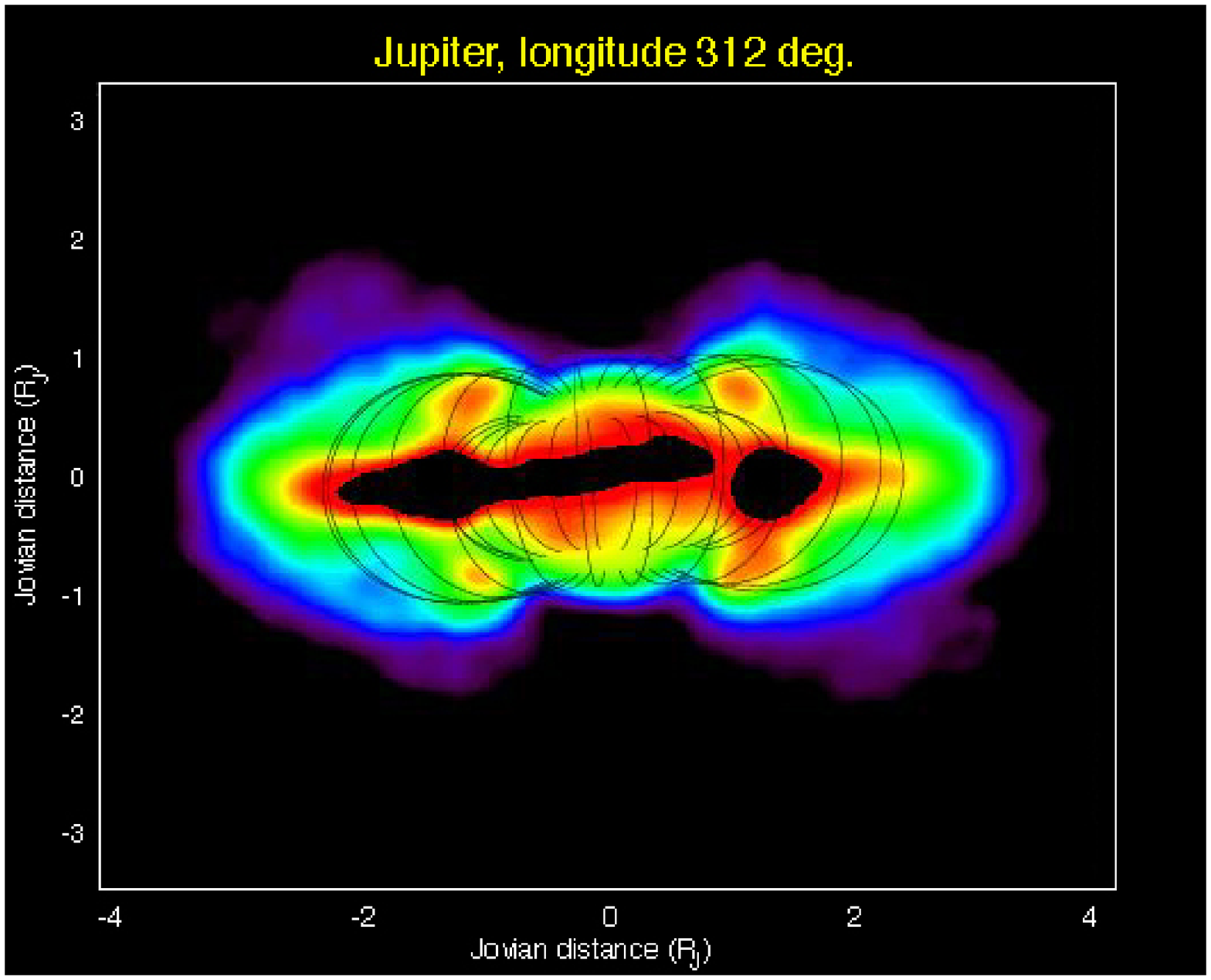}
\end{minipage}
\hspace{0.02\textwidth}
\begin{minipage}[c]{.45\textwidth}
   \centering
   \includegraphics[width=0.9\textwidth]{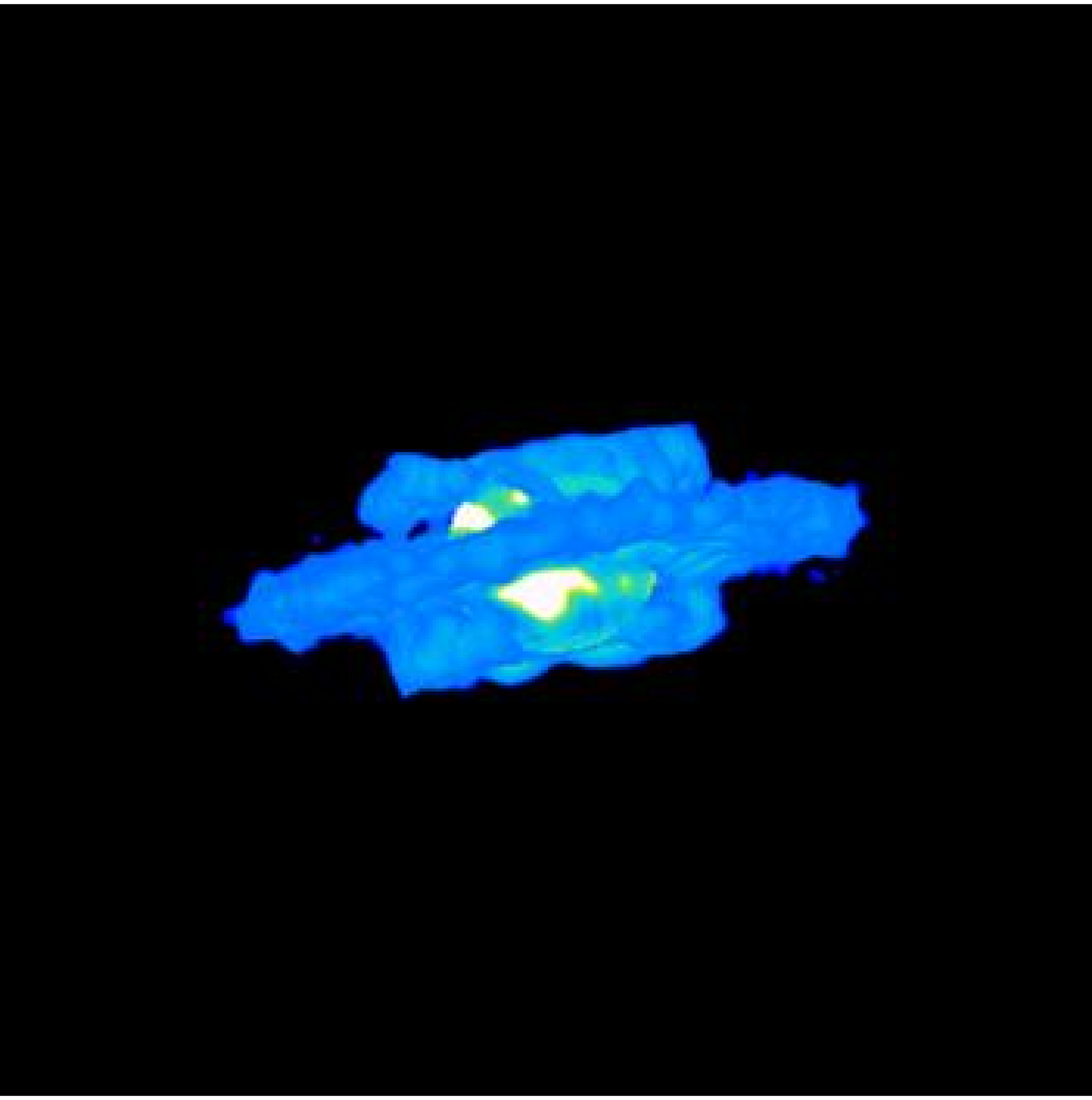}
\end{minipage}
\caption{ Radio images of Jupiter's synchrotron emission.  a) (left) 
   Image made from VLA data taken at a frequency of 1450 MHz.  Both the 
   thermal (confined to Jupiter's disk) and nonthermal emisions are 
   visible.  The resolution is $\sim$0.3 $R_J$, roughly the size of the 
   high latitude emission regions.  Magnetic field lines from a 
   magnetic field model are superposed, shown every $15^\circ$ of 
   longitude.  After \citet{depa97}.  b) (right) Three-dimensional 
   reconstruction of the June 1994 data, as seen from Earth. The planet 
   is added as a white sphere in this visualization.  After 
   \citet{depa98}.
   }
\label{jupsynch}
\end{figure*}

Jupiter's synchrotron radiation has been imaged at frequencies between
74 MHz and 22 GHz 
\citep[see, e.g.,][]{depa91a,depa97,bolt02,depa03a,depa03b}.  A
VLA image of the planet's radio emission at $\lambda=20$ cm is shown
in Figure~\ref{jupsynch}a; the spatial distribution of the synchrotron 
radiation is very similar at all frequencies \citep{depa03b}. 
Because the radio emission is optically thin, and Jupiter rotates in 10 
hours, one can use tomographic techniques to recover the 3D radio 
emissivity, assuming the emissions are stable over 10 hours. An example 
is shown in Figure~\ref{jupsynch}b \citep{saul97,lebl97,depa98}.
The combination of 2D and 3D images is ideal to deduce the particle 
distribution and magnetic field topology from the data 
\citep{dulk97,depa98,dunn03}.

\begin{figure}[tbh]
\centering
\includegraphics[width=0.45\textwidth]{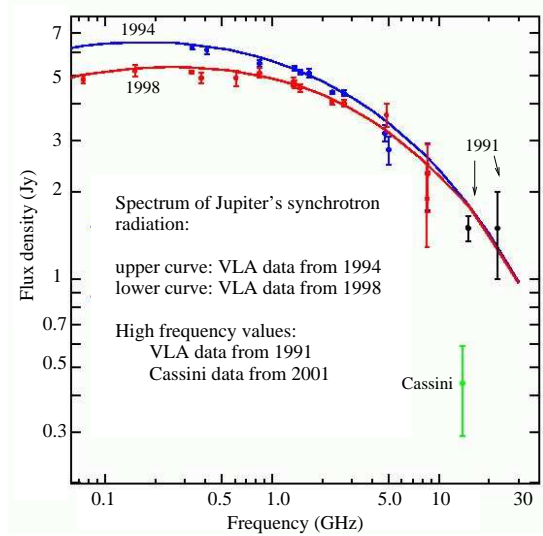}
\caption{ The radio spectrum of Jupiter's synchtrotron emission as 
   measured in September 1998 (lower curve) and June 1994 (upper 
   curve), with high frequency data points from March 1991 (VLA) and 
   January 2001 \citep[Cassini;][]{bolt02}.  Superposed are model 
   calculations that match the data \citep[Adapted from][]{depa03b}.
   }
\label{jupspec}
\end{figure}

The shape of Jupiter's radio spectrum is determined by the intrinsic
spectrum of the synchrotron radiating electrons, the spatial
distribution of the electrons and Jupiter's magnetic field.  Spectra 
from two different years (1994 and 1998) are shown in 
Figure~\ref{jupspec} \citep{depa03c,depa03b}.
The spectrum is relatively flat shortwards of 1-2 GHz, and drops off 
more steeply at higher frequencies. As shown, there are large 
variations over time in the spectrum shortwards of 1-2 GHz, and perhaps 
also at the high frequencies, where the only two existing datapoints at 
15 GHz differ by a factor of $\sim$3. Changes in the radio spectrum 
most likely reflect a change in either the spatial or intrinsic energy 
distribution of the electrons. The large change in spectral shape 
between 1994 and 1998 has been attributed to pitch angle scattering by 
plasma waves, Coulomb scattering and perhaps energy degradation by dust 
in Jupiter's inner radiation belts, processes which affect in 
particular the low energy distribution of the electrons.  With SKA we 
may begin investigating the cause of such variability through its 
imaging capabilities at high angular resolution, and simultaneous good 
u-v coverage at short spacings. As shown by \citet{depa99}, this is 
crucial for intercomparison at different frequencies. With such images 
we can determine the spatial distribution of the energy spectrum of 
electrons, which is tightly coupled to the (still unknown) origin and 
mode of transport (including source/loss terms) of the high energy 
electrons in Jupiter's inner radiation belts.

\subsubsection{Lightning}

Lightning appears to be a common phenomenon in planetary atmospheres.  
It has been observed on Earth, Jupiter, and possibly Venus 
\citep{desc02}.  Electrostatic discharges on Saturn and Uranus have 
been detected by spacecraft at radio wavelengths, and are probably 
caused by lightning.  The basic mechanism for lightning generation in 
planetary atmospheres is believed to be collisional charging of cloud 
droplets followed by gravitational separation of oppositely charged 
small and large particles, so that a vertical potential gradient 
develops.  The amount of charges that can be separated this way is 
limited; once the resulting electric field becomes strong to ionize the 
intervening medium, a rapid 'lightning stroke' or discharge occurs, 
releasing the energy stored in the electric field.  For this process to 
work, the electric field must be large enough, roughly of the order of 
30 V per electron mean free path in the gas, so that an electron gains 
sufficient energy while traversing the medium to cause a collisional 
ionization.  When that condition is met, a free electron will cause an 
ionization at each collision with a gas molecule, producing an 
exponential cascade \citep{gibb99}.

In Earth's atmosphere, lightning is almost always associated with 
precipitation, although significant large scale electrical discharges 
also occur occasionally in connection with volcanic eruptions and 
nuclear explosions.  By analogy, lightning on other planets is only 
expected in atmospheres where both convection and condensation take 
place.  Moreover, the condensed species, such as water droplets, must 
be able to undergo collisional charge exchange.  It is possible that 
lightning on other planets is triggered by active volcanism (such as 
possibly on Venus or Io).

We believe that SKA would be an ideal instrument to search for 
lightning on other planets; the use of multiple beams would facilitate 
discrimination against lightning in our own atmosphere, and 
simultaneous observations at different frequencies would contribute 
spectral information.  For such experiments one needs high time 
resolution (as for pulsars) and the ability to observe over a wide 
frequency range simultaneously, including in particular the very low 
frequencies ($<$ 300 MHz).

\subsection{Thermal emission}

The atmospheres of the giant planets all emit thermal (blackbody) 
radiation. At radio wavelengths most of the atmospheric opacity has 
been attributed to ammonia gas, which has a broad absorption band near 
22 GHz. Other sources of opacity are collision induced absorption by 
hydrogen, H$_2$S, PH$_3$, H$_2$O gases, and possibly clouds. Since the 
overall opacity is dominated by ammonia gas, it decreases approximately 
with $\nu^{-2}$ for $\nu < 22$ GHz. One therefore probes deeper warmer
layers in a planet's atmosphere at lower frequencies.  Spectra of
all four giant planets have been used to extract abundances of
absorbing gases, in particular NH$_3$, and for Uranus and Neptune,
H$_2$S (H$_2$S has been indirectly inferred for Jupiter and Saturn)
\citep[see, e.g.,][]{brig89,depa91b,depa93,debo96}.

\begin{figure}[tbh]
\centering
\includegraphics[width=0.45\textwidth]{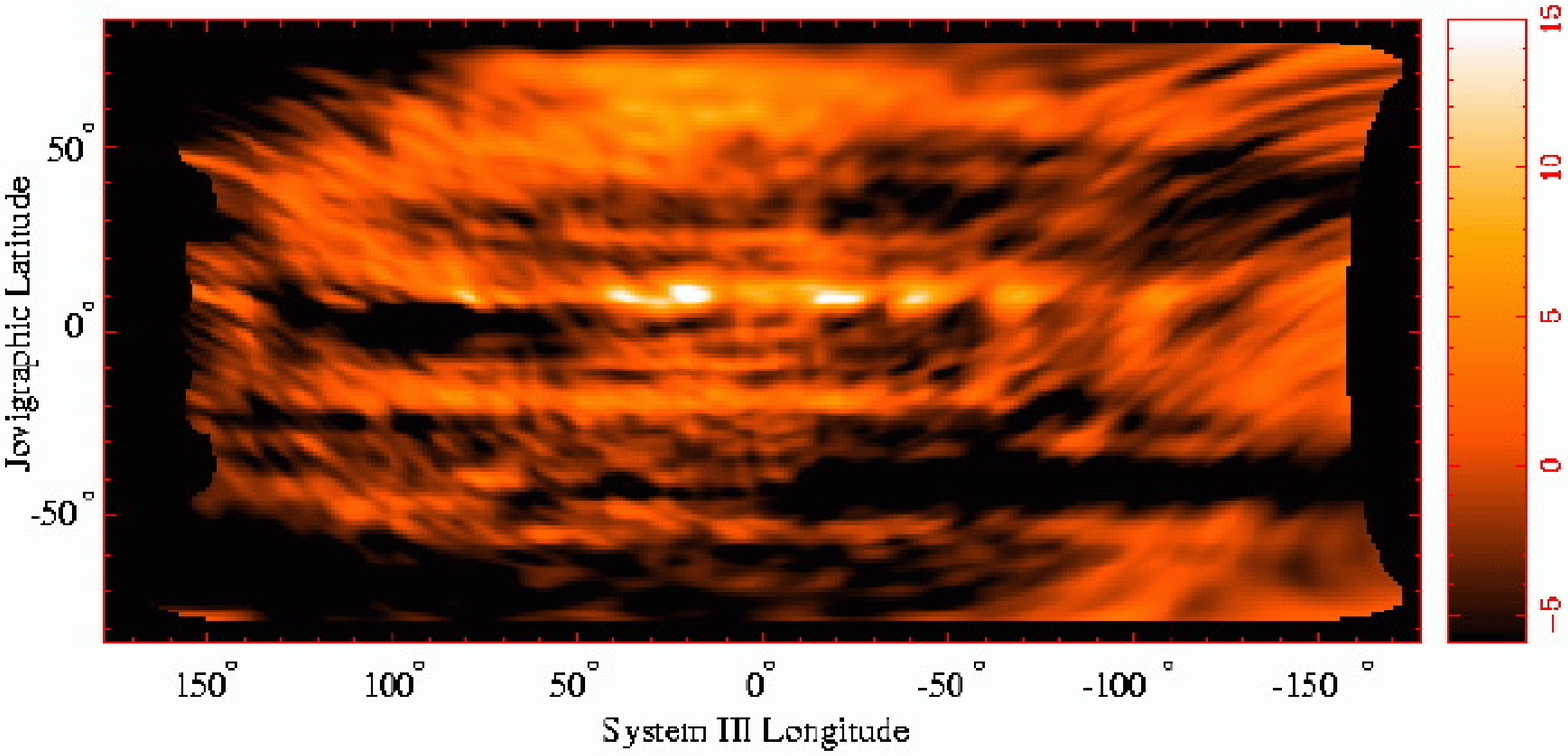}
\caption{Longitude-resolved image of Jupiter at 2 cm \citep{saul04}. }
\label{jup2cm}
\end{figure}

The thermal emission from all four giant planets has been imaged with 
the VLA. To construct high signal-to-noise images, the observations 
need to be integrated over several hours, so that the maps are smeared 
in longitude and only reveal brightness variations in latitude. The
observed variations have typically been attributed to spatial
variations in ammonia gas, as caused by a combination of atmospheric
dynamics and condensation at higher altitudes. Recently, \citet{saul04}
developed an algorithm to construct longitude-resolved images;
they applied this to Jupiter, and their maps reveal, for the first
time, hot spots at radio wavelengths which are strikingly similar to
those seen in the infrared (Figure~\ref{jup2cm}).  At radio wavelengths 
the hot spots indicate a relative absence of NH$_3$ gas, whereas in the 
infrared they suggest a lack of cloud particles. The authors showed 
that the NH$_3$ abundance in hot spots was depleted by a factor of 2 
relative to the average NH$_3$ abundance in the belt region, or a 
factor of 4 compared to zones. Ammonia must be depleted down to 
pressure levels of $\sim$5 bar in the hot spots, the approximate 
altitude of the water cloud.  The algorithm of \citet{saul04} 
only works on short wavelength data of Jupiter, where the synchrotron 
radiation is minimal.

\begin{figure*}[tbh]
\centering
\begin{minipage}[c]{.45\textwidth}
   \centering
   \includegraphics[width=0.9\textwidth]{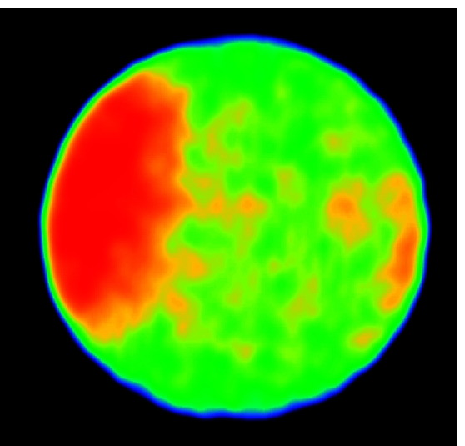}
\end{minipage}
\hspace{0.02\textwidth}
\begin{minipage}[c]{.45\textwidth}
   \centering
   \includegraphics[width=0.9\textwidth]{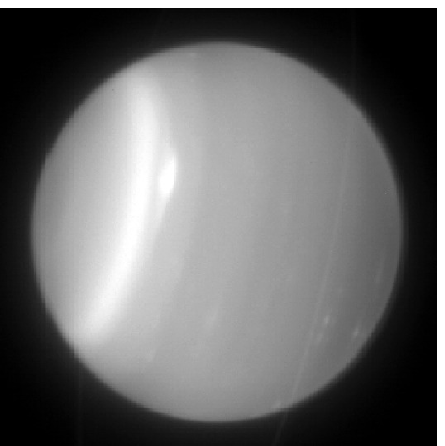}
\end{minipage}
\caption{ Two panels comparing VLA \citep[left,][]{hofs04} and Keck 
   \citep[right,][]{hamm04} images of Uranus from the summer of 2003.  
   In the radio image, red is brighter, cycling to lower brightness 
   through orange, yellow, green, blue, and black.  Note the edge of 
   the radio bright region in the south (to the left in the images) 
   corresponds to a prominent band in the infrared.  The radio bright 
   region in the north has no corresponding band.  The faint line 
   across the planet on the right-hand side of the infrared image is 
   the ring system. 
   }
\label{uranus}
\end{figure*}

Even the longitudinally smeared images are important in deducing the 
state of the deep atmospheres of the giant planets, as attested by 
numerous publications on the giant planets.  Here we discuss 
specifically the case of Uranus, where radio images made with the VLA 
since 1981 at 2 and 6 cm have shown changes in the deep atmosphere 
which appear to be related to the changing insolation as the two poles 
rotate in and out of sunlight over the 40 year uranian year.  Since the 
first images were made, the south pole has appeared brighter than 
equatorial regions.  In the last decade, however, the contrast between 
the two regions and the latitude at which the transition occurs has
changed \citep{hofs03b}.  Figure~\ref{uranus} shows an image from the 
VLA made from data taken in the summer of 2003, along with an image at
near-infrared wavelengths (1.6 $\mu$m) taken with the adaptive optics
system on the Keck telescope in October 2003 \citep{hamm04}.  The VLA
image clearly shows that the south pole is brightest, but it also shows
enhanced brightness in the far-north (to the right on the image).  At 
near-infrared wavelengths Uranus is visible in reflected sunlight, and 
hence the bright regions are indicative of clouds/hazes at high (upper 
troposphere) altitudes, presumably indicative of rising gas (with 
methane condensing out).  We note that the bright band around the south 
pole is at the lower edge of the VLA-bright south polar region.  It 
appears as if air is rising (with condensibles forming clouds) along 
the northern edge of the south polar region and descending over the 
pole, where the low radio opacity is indicative of dry air.

With a sensitivity of SKA 2 orders of magnitude better than that of
the VLA, and excellent instantaneous UV coverage, images of a planet's
thermal emission can be obtained within minutes, rather than hours. 
This would enable direct mapping of hot spots at a variety of 
frequencies, including low frequencies where both thermal and
nonthermal radiation is received.  We can thus obtain spectra of hot
spots, which allow us to derive the altitude distribution of absorbing
gases, something that hitherto could only be obtained via {\it in
situ} probes. Equally exciting is the prospect of constructing
complete 3D maps of the ammonia abundance (or total opacity, to be
precise) at pressure levels between 0.5 and $\sim$ 20-50 bars (these
levels vary some from planet to planet). Will ammonia, and other
sources of opacity, be homogeneous in a planet's deep atmosphere
(i.e., at pressure levels $\gsim$ 10 bar)? Could there be giant
thunderstorms rising up from deep down, bringing up concentrations of
ammonia and other gases from a planet's deep atmosphere, i.e.,
reflecting the true abundance at deep levels? Such scenarios have been
theorized for Jupiter, but never proven \citep{show04}.

A cautionary note here: although excellent images at multiple
wavelengths yield, in principle, information on a giant planet's deep 
atmosphere, detailed modeling will be frustrated in part because of a 
lack of accurate laboratory data on gases and clouds that
absorb at microwave frequencies, such as NH$_3$ and H$_2$O. This 
severely limits the precision at which one can separate contributions 
from different gases. Planetary scientists are in particular eager to
deduce the water abundance in a planet's deep atmosphere (e.g.,
Jupiter).  The potential of deriving the water abundance in the deep 
atmosphere of Jupiter from microwave observations was reviewed by 
\citet{depa04}, while \citet{jans04} investigated the potential of 
using limb darkening measurements on a spinning spacecraft.  These 
studies show that it might be feasible to extract limits on the water 
abundance in the deep atmosphere, but only if the absorption profile of 
water and ammonia gas is accurately known.

\subsection{Rings}

Planetary rings emit thermal radiation, but this contribution is very
small compared to the planet's thermal emission reflected from the 
rings.  Although all 4 giant planets have rings, radio emissions have 
only been detected from Saturn's rings. Other rings are too tenuous to 
reflect detectable amounts of radio emissions (Jupiter's synchrotron 
radiation, though, does reflect the presence of its ring via absorption 
of energetic electrons). Several groups have gathered and analyzed VLA 
data of Saturn's rings over the past decades 
\citep[see, e.g.,][]{gros89,vand99,dunn02}.  These maps, 
at frequencies $>$ a few GHz, are usually integrated over several 
hours, and reveal the classical A, B, and C rings including the Cassini 
Division.  Asymmetries, such as wakes, have been detected in several 
maps; research is ongoing as to correlations between observed 
asymmetries with wavelength and ring inclination angle.

With the high sentivity, angular resolution and simultaneous coverage
of short u-v spacings, maps of Saturn's rings can be improved
considerably. This would allow higher angular resolution and less
longitudinal smearing, allowing searches for longitudinal
inhomogeneities.  In addition, it may become feasible to detect the 
uranian $\epsilon$ ring and perhaps even the main ring of Jupiter 
during ring plane crossings.  We note that the detection of the Jupiter 
ring is made difficult by being so faint and close to an extremely 
bright Jupiter.

\section{Terrestrial Planets}

Radio wavelength observations of the terrestrial planets (Mercury, 
Venus, the Moon, Mars) are important tools for determining 
atmospheric, surface and subsurface properties.  For surface and 
subsurface studies, such observations can help determine temperature, 
layering, thermal and electrical properties, and texture.  For 
atmospheric studies, such observations can help determine temperature, 
composition, and dynamics.  Given the sensitivity and resolution of SKA 
(see Table~\ref{terrresn}), detailed images of both of these types of 
emission will be possible.  We note, however, similarly to the giant 
planet case above, the difficulty in making images with a spatial 
dynamic range of $>$ 10000 (take the case of Venus, with a diameter of 
12000 km, and resolution of $\sim$1 km).  The Moon is a special case, 
where mosaicing will likely be required, the emission is bright and 
complicated, and it is in the near field of SKA (in fact, many of the 
planets are in the near field formally, but the Moon is an extreme 
case).  The VLA has been used to image the Moon \citep{marg97}, 
and near field imaging techniques are being advanced \citep{corn04}, 
but imaging of the Moon will be a challenge for SKA.

\begin{table}[tbh]
\caption{SKA linear resolution for terrestrial planets.}
\label{terrresn}
\vspace*{1truemm}
\footnotesize
\begin{center}
\begin{tabular}{cccc}
\hline\hline
\noalign{\vspace{3pt}}
     & Distance & \multicolumn{2}{c}{resolution (km) $^*$} \\
Body & (AU)     & $\nu=$2 GHz & $\nu=$20 GHz \\
\noalign{\vspace{3pt}}
\hline
\noalign{\vspace{3pt}} 
Moon          & 0.002 & 0.015 & 0.004 \\
Venus         & 0.3   & 2     & 0.7   \\
Mercury, Mars & 0.6   & 4     & 1.3   \\
\noalign{\vspace{3pt}}
\hline\hline
\end{tabular}
\end{center}
\ 

$^*$ assuming maximum baseline of 1000 km
\end{table}

\subsection{Surface and subsurface}

The depth to which temperature variations penetrate in the subsurface 
is characterized by its thermal skin depth, where the magnitude of the 
diurnal temperature variation is decreased by 1/e: \ 
$l_t = \sqrt{k P / (\pi C_p \rho)}$, where $k$ is the thermal 
conductivity, $P$ is the rotational period, $\rho$ is the mass density, 
and $C_p$ is the heat capacity.  For the terrestrial planets, using 
thermal properties of lunar soils and the proper rotation rates, the 
skin depths are of order a few cm (Earth and Mars) to a few 10's of cm 
(Moon, Mercury, and Venus, because of their slow rotation).  The 1/e 
depth to which a radio wavelength observation at wavelength $\lambda$ 
probes in the subsurface is given by: \ $l_r = \lambda / (2 \pi 
\sqrt{\epsilon_r} \tan\Delta)$, where $\epsilon_r$ is the real part of 
the dielectric constant, and $\tan\Delta$ is the ``loss tangent'' of 
the material - the ratio of the imaginary to the real part of the 
dielectric.  For all of the terrestrial planets, given reasonable 
regolith dielectric constant, this is roughly 10 wavelengths.  So, 
the wavelengths of SKA are well matched to probing both above and 
below the thermal skin depths of the terrestrial planets.  

\begin{figure*}[tbh]
\centering
\includegraphics[width=0.9\textwidth]{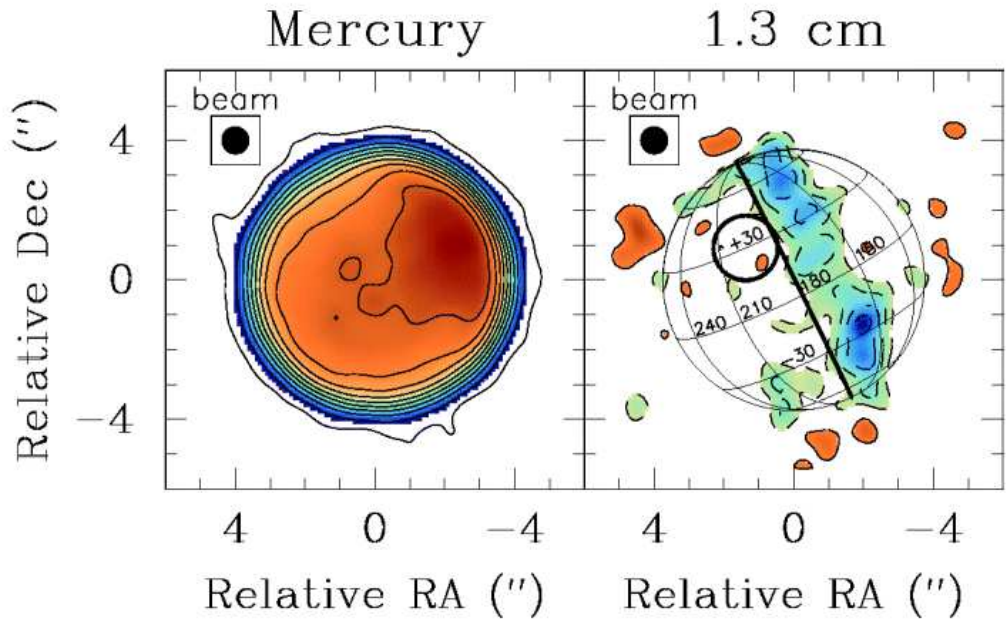}
\caption{ Image of Mercury at 1.3 cm made from data taken at the 
   VLA \citep{mitc94}.  The left panel shows the image, 
   where red is brighter (hotter), cycling to lower brightness through 
   orange, yellow, green, blue, and purple to white.  The right panel 
   shows this image after subtraction of a detailed model.  The solid 
   line is the terminator, the circle is Caloris basin.  The model 
   does well except at the terminator and in polar regions, most likely 
   because of unmodelled topography and surface roughness.
   }
\label{merctherm}
\end{figure*}

The thermal emission from Mercury has been mapped with the VLA and 
BIMA by \citet{mitc94}, who determined that not only was the subsurface 
probably layered, but that the regolith is likely relatively basalt 
free.  Figure~\ref{merctherm} shows a VLA observation, compared with 
the detailed model of Mitchell \& de Pater.  Observations with SKA will 
further determine our knowledge of these subsurface properties.  
Furthermore, given the 1 km resolution, mapping of the near-surface 
temperatures of the polar cold spots \citep[inferred from the presence 
of odd radar scattering behavior - ][]{harm01} will be possible, a 
valuable constraint on their composition.  Finally, given accurate 
enough (well calibrated, on an absolute scale) measurements, 
constraints on the presence or absence of an internal dynamo may be 
placed.

\begin{figure}[tbh]
\centering
\includegraphics[angle=-90,width=0.45\textwidth]{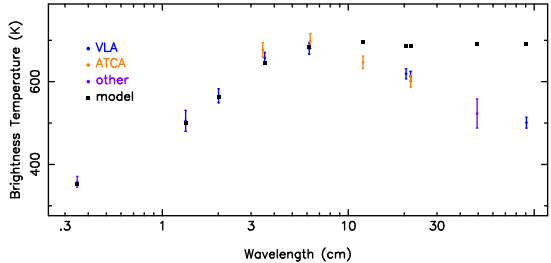}
\caption{ Microwave brightness temperature spectrum of Venus, from 
   \citet{butl03b}.  The depression of the measured emission compared 
   to models at long wavelengths, up to 200 K, is evident.
   }
\label{venus}
\end{figure}

The question of the long wavelength emission from Venus could be 
addressed by SKA observations.  Recent observations have verified 
that the emission from Venus at long wavelengths ($\gsim$ 6 cm) are 
well below predicted - by up to 200 K \citep{butl03b}.
Figure~\ref{venus} shows this graphically.  There is currently no 
explanation for this depression.  Resolved images at long wavelengths 
(say 500 MHz, where the resolution of SKA is of order 100 km at the 
distance of Venus using only the 300 km baselines and less, and the 
brightness temperature sensitivity is about 3 K in 1 hour) will help 
in determining whether this is a global depression, or limited to 
particular regions on the planet.

Although NASA has been sending multiple spacecraft to Mars, there are 
still uses for Earth-based radio wavelength observations.  To our 
knowledge, there is currently no planned microwave mapper for a Mars 
mission, other than the deep sounding very long wavelength radar 
mappers (MARSIS, for example).  So observations in the meter-to-cm 
wavelength range are still important for deducing the properties of the 
important near-surface layers of the planet.  Observations of the 
seasonal caps as they form and subsequently recede would provide 
valuable constraints on their structure.  Observations of the odd 
``stealth'' region \citep{edge97} would help constrain its composition 
and structure, and in combination with imagery constrain its 
emplacement history.

\subsection{Atmosphere}

The Moon and Mercury have no atmosphere to speak of, but Venus and Mars 
will both benefit from SKA observations of their atmospheres.  
Short wavelength observations of the venusian atmosphere ($\lsim$ 3 cm) 
probe the lower atmosphere, below the cloud layer ($\lsim$ 40 km).  
Given the abundance of sulfur-bearing molecules in the atmosphere, and 
their high microwave opacity, such observations can be used to 
determine the abundances and spatial distribution of these molecules.  
\citet{jenk02} have mapped Venus with the VLA at 1.3 and 2 cm, 
determining that the below-cloud abundance of SO$_2$ is lower than that 
inferred from infrared observations, and that polar regions have a 
higher abundance of H$_2$SO$_4$ vapor than equatorial regions, 
supporting the hypothesis of Hadley cell circulation.  VLA observations 
are hampered both by sensitivity and spatial dynamic range.  The EVLA 
will solve part of the sensitivity problem, but will not solve the 
instantaneous spatial dynamic range problem - only the SKA can do both. 
Given SKA observations, cloud features (including at very small 
scales), and temporal variation of composition (which could be used as 
as proxy to infer active volcanism, since it is thought that 
significant amounts of sulfur-bearing molecules would be released in 
such events) could be sensed and monitored.

\begin{figure}[tbh]
\centering
\includegraphics[width=0.45\textwidth]{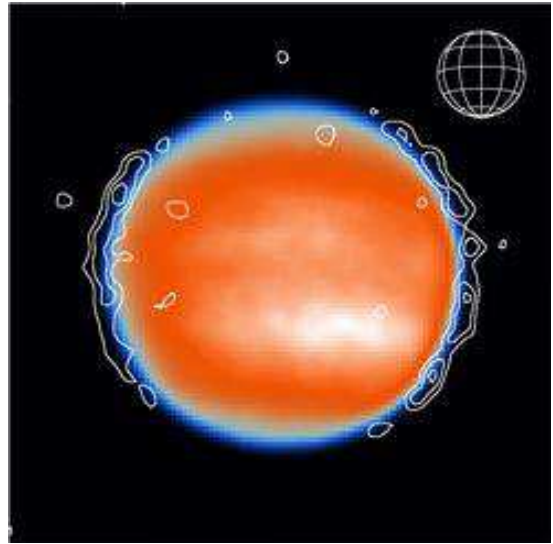}
\caption{ Map of water vapor in the Mars atmosphere made from data 
   taken at the VLA in 1991.  The colored background is the thermal 
   emission from the surface.  The contours are the H$_2$O emission, 
   seen only along the limb.  From \citet{clan92}.
   }
\label{marsh2o}
\end{figure}

Observations of the water in the Mars atmosphere with the VLA have 
provided important constraints on atmospheric conditions and the 
climate of the planet \citep{clan92}.  The 22 GHz H$_2$O line is 
measured, and emission is seen along the limb, where pathlengths are 
long (this fact is key - the resolution of the atmosphere along the 
limb is critical).  Figure~\ref{marsh2o} shows an image of this.  For 
added sensitivity in these kinds of observations (needed to improve the 
deduction of temperature and water abundance in the atmosphere), only 
the SKA will help.

\section{Large Icy Bodies}

In addition to their odd radar scattering properties (see the Radar 
section below), the Galilean satellites Europa and Ganymede exhibit 
unusually low microwave emission \citep{depa84,muhl86,muhl91b}.  
Observations with SKA will determine the deeper subsurface properties 
of the Galilean satellites, Titan, the larger uranian satellites, and 
even Triton, Pluto, and Charon.  For example, given a resolution of 40 
km at 20 GHz (appropriate for the mean distance to Uranus), maps of 
hundreds of pixels could be made of the uranian moons Titania, Oberon, 
Umbriel, Ariel, and Miranda.  Pushing to 3000 km baselines, maps of 
tens of pixels could even be made of the newly discovered large KBOs 
Quaoar and Sedna (Quaoar is estimated to be $\sim$40 masec in diameter, 
Sedna about half that, \citep{brow04a,brow04b}).  These bodies are some 
10's of K in physical temperature, probably with an emissivity of 
$\sim$0.9 (by analog with the icy satellites), so with a brightness 
temperature sensitivity of a few K in a few masec beam, SKA should have 
no problem making such maps with an SNR of the order of 10's in each 
pixel.  SKA will be unique in its ability to make such maps of these 
bodies - optical images will come nowhere near this resolution unless 
space-based interferometers become a reality.

\section{Small Bodies}

Perhaps the most interesting solar system science with SKA will involve 
the smaller bodies in the solar system.  Because of their small size, 
their emission is weak, and they have therefore not been studied very 
extensively, particularly at longer wavelengths.  Such bodies include 
the smaller satellites, asteroids, Kuiper Belt Objects (KBOs), and 
comets.  These bodies are all important probes of solar system 
formation, and will yield clues as to the physical and chemical state 
of the protoplanetary and early planetary environment, both in the 
inner and outer parts of the solar system.

\subsection{Small satellites}

It is sometimes hypothesized that Phobos and/or Deimos are captured 
asteroids because of surface spectral reflectivity properties.  This 
is inconsistent, however, with their current dynamical state and 
low internal density 
\citep[see, e.g., the discussions in][]{burn92,rivk02}.  The two moons 
could also have been formed via impact of a large asteroid into Mars, 
which could also have helped in forming the north-south dichotomy on 
the planet \citep{crad94}.  Determination of the properties of the 
surface and near-surface could help unravel this mystery.  These bodies 
are $\sim$10 km in diameter, so at opposition will be $\sim$30 masec in 
apparent diameter, so SKA will be able to map them with a few 10's of 
pixels on the moons.  This will provide some of these important 
properties and their variation as a function of location on the moons 
(notably regolith depth and thermal and electrical properties).  As 
another example, consider the eight outer small jovian satellites, 
about which little is known, either physically or chemically.  All 
eight of them, with diameters of from 15 to 180 km (Himalia), could be 
resolved by SKA at 20 GHz, determining their shapes as well as their 
surface and subsurface properties.

We note, however, that the imaging of these small satellites can be 
challenging, as they are often in close proximity to a very bright 
primary which may have complex brightness structure.  As such, even 
with the specification that SKA must have a dynamic range of 10$^6$, 
it will not be trivial to make images of these small, relatively 
weak satellites.  

\subsection{Main Belt Asteroids}

The larger of the main belt asteroids 
are the only remaining rocky protoplanets (bodies of order a few 
hundred to 1000 km in diameter), the others having been dispersed or 
catastrophically disrupted, leaving the comminuted remnants comprising 
the asteroid belt today \citep{davi79}.  They have experienced 
divergent evolutionary paths, probably as a consequence of forming on 
either side of an early solar system dew line beyond which water was a 
significant component of the forming bodies.  Vesta is thought to have 
accreted dry, consequently experiencing melting, core formation, and 
volcanism covering its surface with basalt \citep{drak01}.  Ceres and 
Pallas, thermally buffered by water never exceeding 400K, experienced 
aqueous alteration processes evidenced by clay minerals on their 
surfaces \citep{rivk97}.  These three large MBAs all reach apparent 
sizes of nearly 1\arcsec\ at opposition, so maps with hundreds of 
pixels across them can be made, with high SNR (brightness temperature 
is of order 200 K, while brightness temperature sensitivity is of order 
a few K).  Such maps will directly probe regolith depth and properties 
across the asteroids, yielding important constraints on formation 
hypotheses.

SKA will also be able to detect and map the smaller MBAs.  Given the 
distances of the MBAs to the Sun, they typically have 
surface/sub-surface brightness temperatures (the brightness temperature 
is just the physical temperature multiplied by the emissivity) of 
$\sim$ 200 K.  Given a typical distance (at opposition) of 1.5 AU, 
this gives diameters of 2, 20, and 200 masec for MBAs of 1, 10, and 
100 km radius, with flux densities of 0.3, 30, and 3000 $\mu$Jy at 
$\lambda$ = 1 cm.  So the larger MBAs will be trivial to detect and 
map, but the smallest of them will be somewhat more difficult to 
observe (but not beyond the sensitivity of SKA - see the discussion 
above on Instrumental Capabilities).  There are more than 1500 MBAs 
with diameter $>$ 20 km just in the IRAS survey \citep{tede02}.

\subsection{Near Earth Asteroids}

In addition to being important remnants of solar system formation, NEAs 
are potential hazards to us here on Earth \citep{morb02}.  As such, 
their characterization is important \citep{cell02}.  SKA will easily 
detect and image such asteroids.  As they pass near the Earth, they are 
typically at a brightness temperature of 300 K, and pass at a distance 
of a few lunar radii ($\sim$0.005 AU).  This distance gives diameters 
of 6, 60, and 600 masec for NEAs of 10, 100, and 1000 m, with flux 
densities of 0.005, 0.5, and 50 mJy $\mu$Jy at $\lambda$ = 1 cm.  
Again, these will be easily detected and mapped.  ALMA will also be an 
important instrument for obseving these bodies \citep{butl01}, but it 
is the combination of the data from ALMA and SKA that allows a complete 
picture of the surface and subsurface properties to be formed.

\subsection{Kuiper Belt Objects}

In general KBOs are detected at optical/near-IR wavelengths in 
reflected sunlight.  Since the albedo of comet Halley was measured (by 
spacecraft) to be 0.04, comets/KBOs are usually assumed to have a 
similar albedo (which most likely is not true).  This assumed albedo is 
then used to derive an estimate of the size based on the magnitude of 
the reflected sunlight.  Resolved images, and hence size estimates, 
only exist of the largest KBOs \citep[see, e.g.,][]{brow04a}.  The only 
other possibility \citep[ignoring occultation experiments -][]{coor03}
to determine the size is via the use of radiometry, where observations 
of both the reflected sunlight and longer wavelength observations of 
thermal emission are used to derive both the albedo and radius of the 
object.  This technique has been used, for example, for asteroids in 
the IRAS sample \citep{tede02}.  Although more than 100 KBOs have been 
found to date, only two have been detected in direct thermal emission, 
at wavelengths around 1 mm \citep{jewi01,marg02a} - the emission is 
simply too weak.  ALMA will be an extremely important telescope for 
observing KBOs \citep{butl01}, but will just barely be able to resolve 
the largest KBOs (with a resolution of 12 masec at 350 GHz in its most 
spread out configuration).  SKA, with a resolution of a few masec and a 
brightness temperature sensitivity of a few K, will resolve all of the 
larger of the KBOs (larger than 100 km or so), and will easily detect 
KBOs with radii of 10's of km.  Combined observations with ALMA and SKA 
will give a complete picture of the surface, near-surface, and deeper 
subsurface of these bodies.

\subsection{Comets}

In addition to holding information on solar system formation, comets 
are also potentially the bodies which delivered the building blocks of 
life (both simple and complex organic molecules) to Earth.  As such, 
they are important astronomical targets, as we would like to understand 
their current properties and how that constrains their history.

\subsubsection{Nucleus}

Long wavelengths (cm) are nearly unique in their ability to probe right 
to the surfaces of active comets.  Once comets come in to the inner 
solar system, they generally produce so much dust and gas that the 
nucleus is obscured to optical, IR, and even mm wavelengths.  At cm 
wavelengths, however, one can probe right to (and into) the nucleus of 
all but the most productive comets.  For example, comet Hale-Bopp was 
detected with the VLA at X-band \citep{fern02}.  Given nucleus sizes of 
a few to a few 10's of km, and distances of a few tenths to 1 AU, the 
flux densities from cometary nuclei should be from about 1 $\mu$Jy to 1 
mJy at 25 GHz - easy to detect with SKA.  Multi-wavelength observations 
should tell us not only what the surface and near-surface density is, 
but if (and how) it varies with depth.  These nuclei should be roughly 
10-100 masec in apparent size, so can be resolved at the high 
frequencies of SKA.  With resolved images, in principle it would be 
possible to determine which areas were covered with active (volatile) 
material, i.e., ice, and which were covered with rocky material, and 
for the rocky material whether it was dust (regolith) or solid rock.  

\subsubsection{Ice and dust grain halo}

Large particles (rocks and ice cubes) are clearly shed from cometary 
nuclei as they become active, as shown by radar observations 
\citep{harm99}.  The properties of these activity byproducts are 
important as they contain information on the physical structure and 
composition of the comets from which they are ejected.  Observations 
at the highest SKA frequencies should be sensitive to emission from 
these large particles (even though they also probe down through them 
to the nucleus), and can thus be used to make images of these 
particles - telling us what the distribution (both spatially, and the 
size distribution of the particles) and total mass is, and how it 
varies with time.

\subsubsection{Coma}

Observations of cometary comae will tell us just what the composition 
of the comets is - both the gas to dust ratio, and the relative ratios 
of the volatile species.  Historically, observations of cometary 
comae at cm wavelengths have been limited to OH, but with the 
sensitivity of SKA, other molecules such as formaldehyde 
\citep[detected in comet Halley with the VLA - ][]{snyd89} and CH 
should be observable.  The advantage of long wavelength transitions is 
that we observe rotational transitions of the molecules, which are much 
easier to understand and accurately characterize in the statistical 
equilibrium and radiative transfer models (necessary to turn the 
observed intensities into molecular abundances).  Millimeter wavelength 
observations of cometary molecules have proven fertile ground 
\citep[see, e.g.,][]{bive02}, but the cm transitions of molecules are 
also important as they probe the most populous energy states, and some 
unique molecules which do not have observable transitions in the 
mm-submm wavelengths.

The volatile component of comets is $\sim$80\% water ice, with the bulk 
of the rest CO$_2$.  All other species are present only in small 
quantities.  As the comet approaches the Sun, the water starts to 
sublimate, and along with the liberated dust forms the coma and tails.  
At 1 AU heliocentric distance, the typical escape velocity of the water 
molecules is 1 km/s, and the lifetime against dissociation is about 
80000 sec, which leads to a water coma of radius 80000 km.  Although 
there have been some claims of direct detection of the 22 GHz water 
line in comets, a very sensitive search for this emission from 
Hale-Bopp detected no such emission \citep{grah00}.  With SKA, such 
observations should be possible and will likely be attempted.  The 
problem is that the resolution is too high - with such a large coma, 
most of the emission will be resolved out.  

\begin{figure*}[tbh]
\centering
\includegraphics[width=0.8\textwidth]{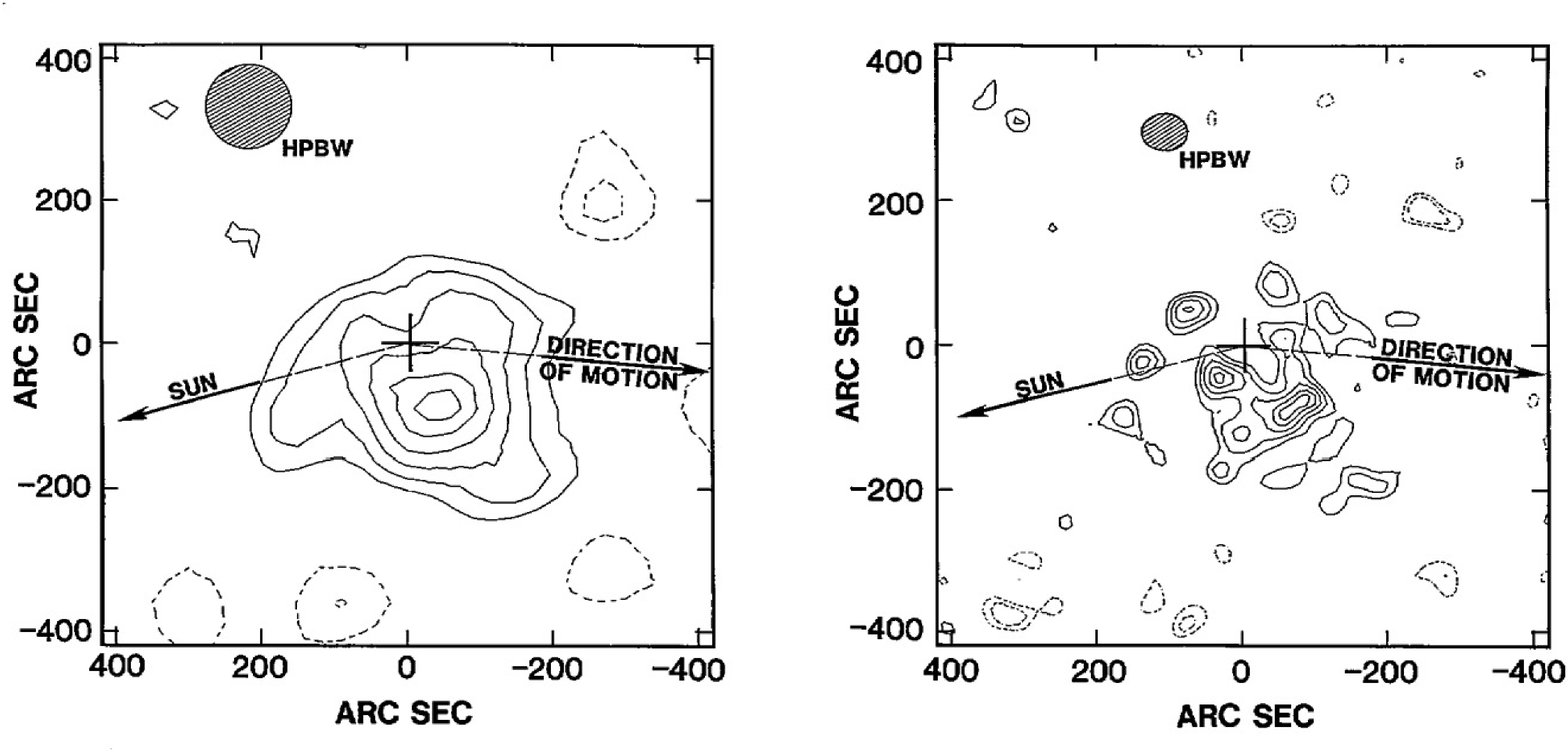}
\caption{ Images of the OH emission from comet Halley made with the 
   VLA at low (left) and high (right) resolution.  From \citet{depa91c}.
   }
\label{halley}
\end{figure*}

\begin{figure}[tbh]
\centering
\includegraphics[width=0.45\textwidth]{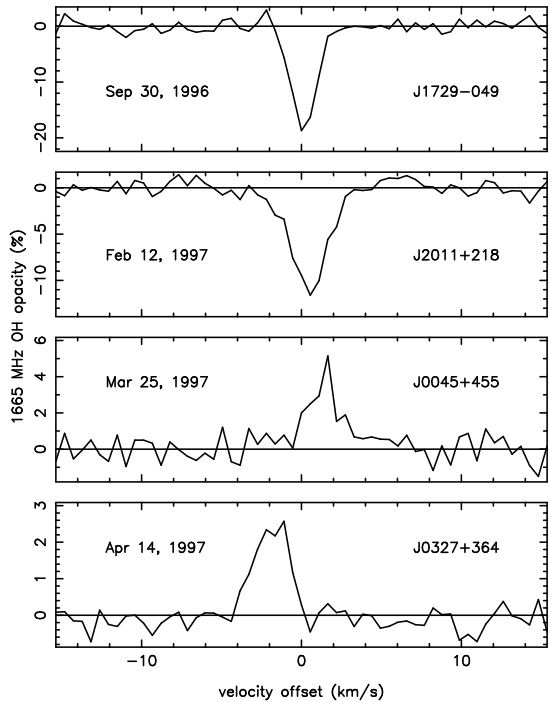}
\caption{ Spectra of the OH emission from comet Hale-Bopp made as the 
   comet occulted background sources.  From \citet{butl97b}.
   }
\label{halebopp}
\end{figure}

Most of the water dissociates into H and OH.  The hydroxyl has a 
lifetime of $1.6 \times 10^5$ sec at 1 AU heliocentric distance, 
implying a large OH coma - of order 10\arcmin\ apparent size at 1 AU 
geocentric distance.  The OH is pumped into disequilibrium by 
solar radiation, and acts as a maser.  As such, the emission can be 
quite bright, and is regularly observed at cm wavelengths by single 
dishes as it amplifies the galactic or cosmic background \citep{schl85}.
Since the spatial scale is so large, however, VLA observations of 
cometary OH have been limited to observations of only a few 
comets - Halley, Wilson, SL-2, and Hale-Bopp \citep{depa91c,butl97b}.
Figure~\ref{halley} shows a VLA image of the OH emission from comet 
Halley.  Though scant, these observations have helped demonstrate that 
the OH in cometary comae is irregularly distributed, likely due to 
quenching of the population inversion from collisions in the inner 
coma.  Similar to the case for water above, SKA will resolve out most 
of this emission.  It will certainly provide valuable observations of 
the distribution of OH in the inner coma, but not much better than is 
possible with the VLA currently.  The real power of the SKA will be in 
observations of background sources amplified by the OH in the coma.  
The technique is described in \citet{butl97a} and was demonstrated 
successfully for Hale-Bopp \citep{butl97b}.  Figure~\ref{halebopp} 
shows example spectra.  As a comet moves relative to a background 
source, the OH abundance along a chord through the coma is probed.  
Since the SKA will be sensitive to very weak background sources, many 
such sources should be available for tracking at any time, providing a 
nearly full 2-D map of the coma at high resolution (each chord is 
sampled along a pencil beam through the coma with diameter 
corresponding to the resolution of the interferometer at the distance 
of the comet).  Combined with single-dish observations, this should 
provide a very accurate picture of the OH in cometary comae.

Among the five most common elements in cometary comae, the chemistry 
involving nitrogen is one of the least well understood (along with 
sulfur).  In addition, ammonia is particularly important in terms of 
organic precursor molecules, and can also be used as a good thermometer 
for the location where comets formed - whether the nitrogen is in N$_2$ 
or NH$_3$ depends on the temperature of the local medium, among other 
things \citep{char02}.  Ammonia has a rich microwave spectrum which has 
been extensively observed in interstellar molecular clouds 
\citep[see, e.g.,][]{ho83}.  Observations of ammonia in cometary comae 
are therefore potentially very valuable in terms of determining current 
chemistry and formation history.  Recently, comets Hyakutake and 
Hale-Bopp were observed in NH$_3$ \citep{bird97,hiro99,butl02}.  
Observations of cometary NH$_3$ will suffer from the same problem as 
the H$_2$O and OH - the NH$_3$ coma is large (although about a factor 
of 10 smaller than the water coma).  However, if the individual 
elements are relatively small, and have any reasonable single dish 
capability, the NH$_3$ may still be detected.

\section{Radar}

Radar observations of solar system bodies contribute significantly to
our understanding of the solar system.  Radar has the potential to
deliver information on the spin and orbit state, and the surface and
subsurface electrical properties and texture of these bodies.  The two
most powerful current planetary radars are the 13 cm wavelength system
on the 305 m Arecibo telescope and the 3.5 cm system on the 70 m
Goldstone antenna.  A radar that made use of the SKA for both
transmitting and receiving the echo would have a sensitivity many
hundreds of times greater than the Arecibo system, the most sensitive
of the two current systems.  However,  while, in theory, it would be
possible to transmit with all, or a substantial fraction of, the SKA
antennas, the additional complexity of controlling transmitters at each
antenna, providing adequate power and solving atmospheric phase
problems makes this option potentially prohibitively expensive.  Used
with the Arecibo antenna as a transmitting site, an Arecibo/SKA radar
system would have 30 to 40 times the sensitivity of the current Arecibo
planetary radar accounting for integration time and possible use of a
shorter wavelength than 13 cm.  If it were combined with a specially
built transmitting station (100 m antenna equivalent size, 5 MW of
transmitted power, 3 cm wavelength) the SKA would have 150 to 200 times
the sensitivity of the current Arecibo system.  This sensitivity would
open up new areas of solar system studies especially those related to
small bodies and the satellites of the outer planets. 

\begin{figure}[tbh]
\centering
\includegraphics[width=0.45\textwidth]{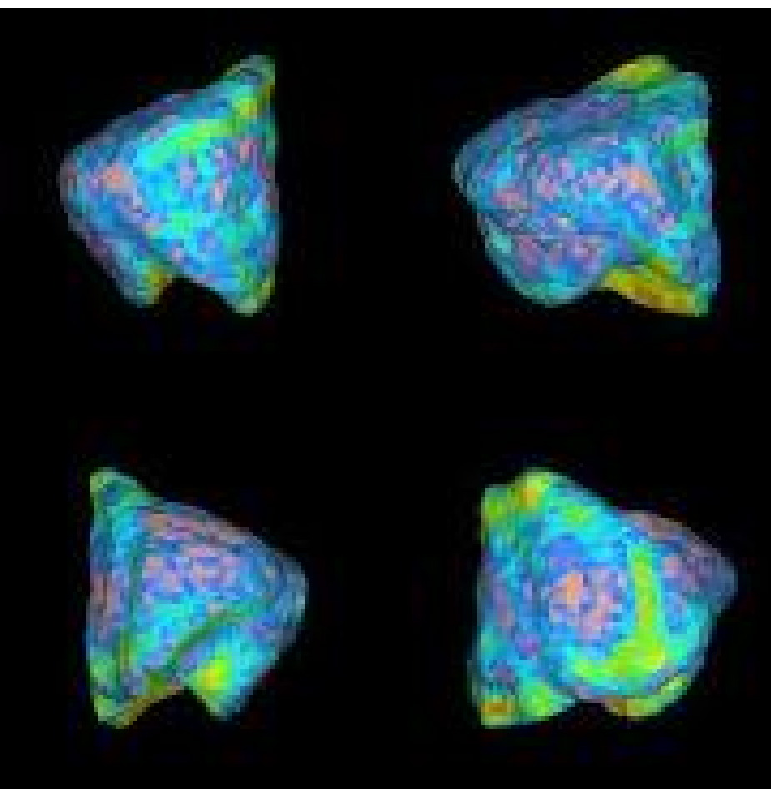}
\caption{ 
   A shape model for the 0.5 km NEA 6489 Golevka derived from Arecibo 
   delay-Doppler images \citep{huds00,ches03}.  The colors indicate the 
   relative size of local gravitational slopes.  
   }
\label{golevka}
\end{figure}

\begin{figure*}[tbh]
\centering
\begin{minipage}[c]{.45\textwidth}
   \centering
   \includegraphics[width=0.9\textwidth]{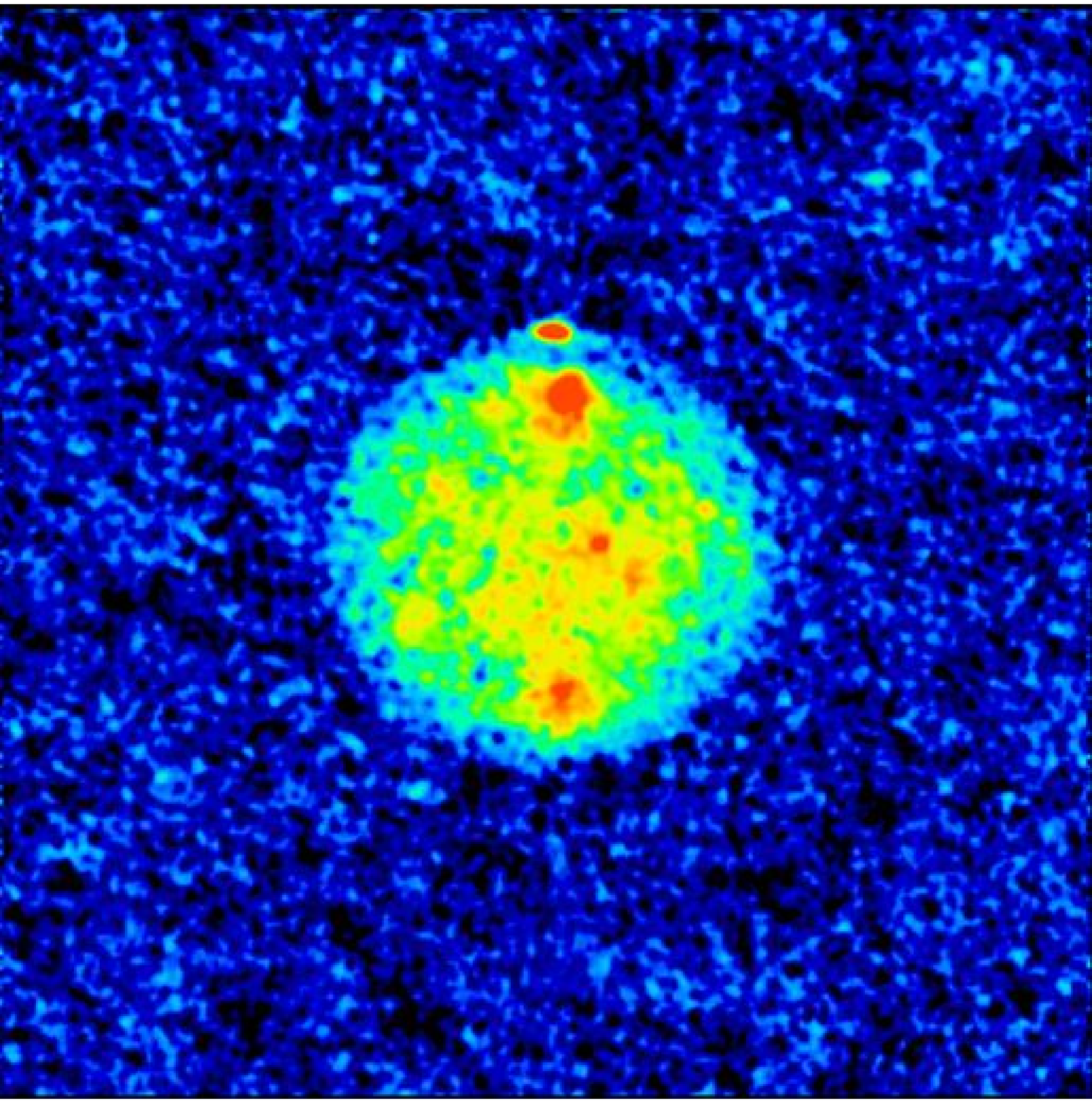}
\end{minipage}
\hspace{0.02\textwidth}
\begin{minipage}[c]{.45\textwidth}
   \centering
   \includegraphics[width=0.9\textwidth]{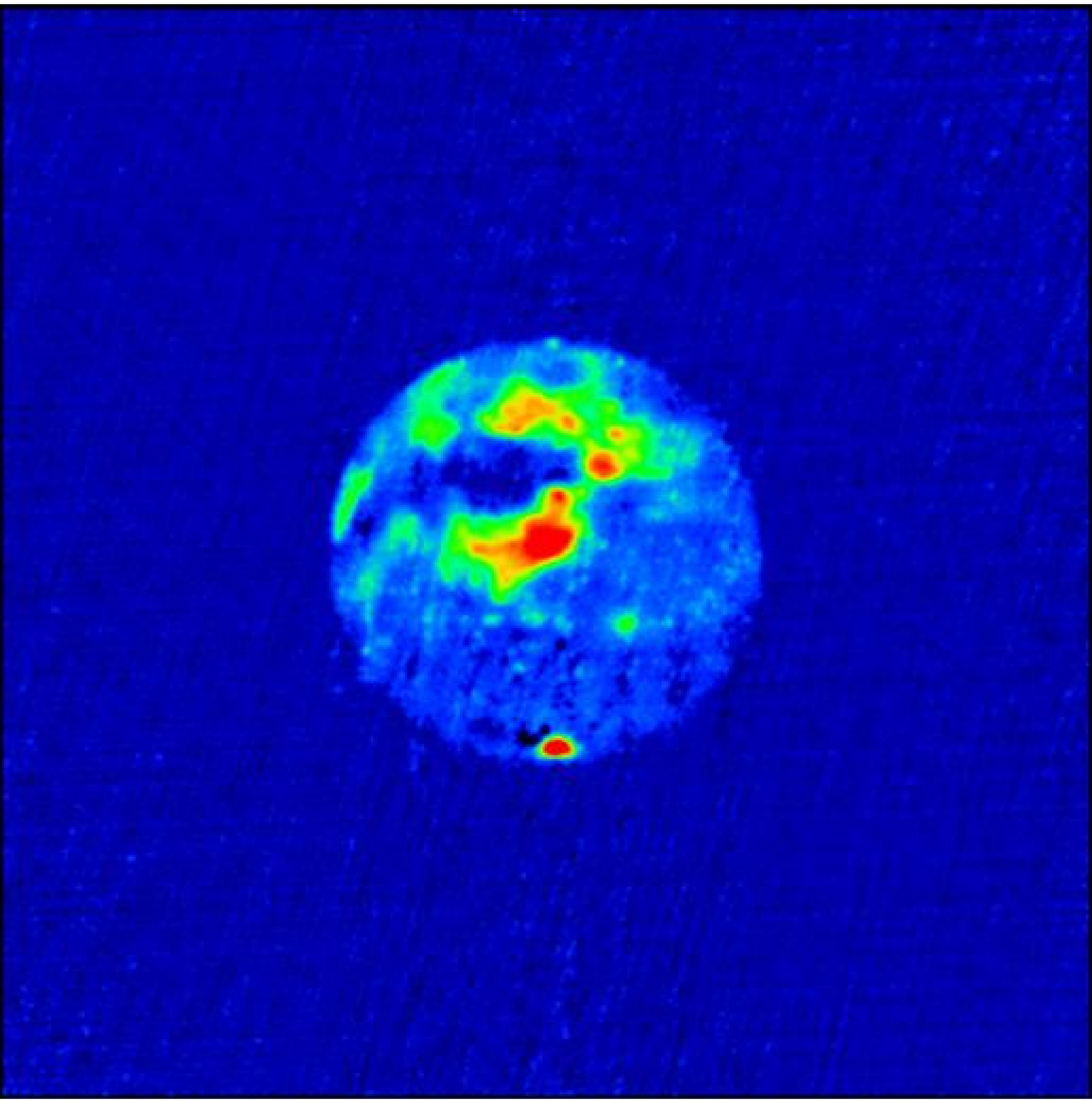}
\end{minipage}
\caption{ Images made with the combined Goldstone+VLA radar instrument. 
   Mars (left) observations done in October 1988.  Mercury (right) 
   observations done in August 1991.  In both images, areas of brighter 
   radar reflectivity are red, cycling to lower reflectivity through 
   orange, yellow, green, light blue, blue, purple, and black.  After 
   \citep{muhl91a,butl93,muhl95}.
   }
\label{gvla}
\end{figure*}

Imaging with the current planetary radar systems is achieved by either
measuring echo power as a function of the target body's delay
dispersion and rotationally induced Doppler shift - delay-Doppler
mapping - or by using a radio astronomy synthesis interferometer system
to spatially resolve a radar illuminated target body.  Delay-Doppler
mapping of nearby objects such a Near Earth asteroids (NEAs) can
achieve resolutions as high as 15 m (Figure~\ref{golevka}) but such
images suffer from ambiguity (aliasing) problems due to two or more
locations on the body having the same distance and velocity relative to
the radar system.  Synthesis imaging of radar illuminated targets
provides unambiguous plane-of-sky images but, to date, the spatial
resolution has been considerably less than can be achieved by
delay-Doppler imaging.  A noted example of the synthesis imaging
technique was the discovery of water ice at the poles of Mercury by
using the Goldstone transmitter in combination with the Very Large
Array (VLA), another is the discovery of the so-called ``Stealth''
region on Mars by that same combined radar (Figure~\ref{gvla}).  As
discussed below, using the SKA as a synthesis instrument will not
provide adequate spatial resolution for studies of Near Earth Objects
(NEOs) but it will resolve them, mitigating the effects of ambiguities
in delay-Doppler imaging.

\subsection{Terrestrial planets}

At the distances of the closest approaches of Mercury, Venus and Mars
to the Earth, the spatial resolution of a 3,000 km baseline SKA at 10
GHz will be approximately 1 km, 0.5 km and 0.7 km, respectively. The
SKA-based radar system would be capable of imaging the surface of
Mercury at 1 km resolution with a 1.0-sigma sensitivity limit
corresponding to a radar cross section per unit area of about -30 db,
good enough to map to very high incidence angles. For Mars, the
equivalent spatial resolution for the same sensitivity limit would be
$<$ 1 km. The very high absorption in the Venus atmosphere at 10 GHz
would reduce the echo strength and, hence, limit the achievable
resolution.  However, short wavelength observations would complement
the longer 13 cm imagery from Magellan, provide additional information
about the electrical properties of the surface via studies of the
polarization properties of the echo \citep{hald97,cart04}, and monitor
the surface for signs of current volcanic activity. For both Mercury
and Mars, radar images at 1 km resolution would potentially be of great
interest for studying regolith properties on Mercury and probing the
dust that covers much of the surface of Mars. For the polar ice
deposits on Mercury the sensitivity would allow sub-km resolution,
significantly better than the 2 km Arecibo delay-Doppler imagery of
\citet{harm01}.  However, this will require the capability to perform 
delay-Doppler imaging within the SKA's synthesized beam areas. 

\subsection{Icy Satellites}

Radar is uniquely suited to the study of icy surfaces in the solar
system and a SKA based system would provide images (or at least
detections) of these bodies in the parameters responsible for their
unusual radar scattering properties. As shown by recent Arecibo radar
observations of Iapetus, the third largest moon of Saturn, the radar
reflection properties of icy bodies can be used to infer surface
chemistry in that pure ice surfaces can be distinguished from ones
which in- corporate impurities such as ammonia that suppresses the low
loss volume scattering properties of the ice \citep{blac04} The unusual
radar scattering properties of the Galilean satellites have been known
for some time \citep{camp78,ostr92}.  As such, they are inviting
targets for a SKA radar. At a distance to the jovian system of 4.2 AU,
the smallest spatial size of the SKAs synthesized beam would be about 6
km while, given the very high backscatter cross sections of the icy
Galilean satellites, signal-to-noise considerations would allow imaging
with about 5 km resolution, a good match to the size of the synthesized
beam. Depending on the prospects for NASAs proposed Jupiter Icy Moons
Mission (JIMO) and its instrument payload, radar images of the icy
moons at resolutions of a few km would provide unique information about
the regoliths/upper surface layers of the icy satellites. Past radar
observations of Titan have been instrumental in shaping our ideas of
what resides on the surface there - the existence of a deep, global
methane/ethane ocean was disproved \citep{muhl90}, but recent Arecibo
radar observations have provided evidence for the possible presence of
small lakes or seas \citep{camp03} The Cassini mission, just arriving
in the saturnian system, will make radar reflectivity measurements of
Titan, but they will not be global, nor will the resolution be as fine
as desired. At a distance of 8.0 AU, the spatial resolution of a 3,000
km baseline SKA at 10 GHz will be approximately 12 km - global radar
imagery at this scale would be a powerful tool for studying the surface
and subsurface of this enigmatic body. Given the extreme sensitivity of
the SKA for radar observations, it would even be possible to make
detections of Triton and Pluto. At the distances of these bodies, it
will probably not be possible to make resolved images of them (although
theoretically it is possible, given the SKA resolution) we can still at
least measure the bulk properties of their surfaces and make crude
hemispherical maps.

A SKA system could also be used to investigate the radar scattering
properties of some of the smaller satellites of Jupiter, Saturn and
Uranus. It will be possible to investigate the radio wavelength
scattering properties of most of the satellites of Jupiter, satellites
of Saturn with larger than 50 to 100 km and the five large satellites
of Uranus.

\subsection{Small bodies}

\subsubsection{Primary scientific objectives}

While spacecraft have imaged a small number of asteroids and comets,
Earth based planetary radars will be the dominant means for the
foreseeable future for obtaining astrometry, and determining the
dynamical state and physical properties of small bodies in the inner
solar system.  Internal structure and collisional histories, important
for solar system formation theories, can be deduced from measurements
of asteroid sizes and shapes and from detailed imagery of their
surfaces. Variations in the reflection properties of main belt
asteroids with distance from the sun could pinpoint the transition
region from rocky to icy bodies, again important for theories of solar
system formation. There is also considerable uncertainty as to the size
distribution of comets that a SKA based radar system could resolve.
\citet{bern04} have pointed out that there is a significant shortage of 
KBOs at small sizes if comets have nuclei that are in the 10 km range 
as currently thought.

\subsubsection{Near Earth Asteroids}

\begin{figure}[tbh]
\centering
\includegraphics[width=0.45\textwidth]{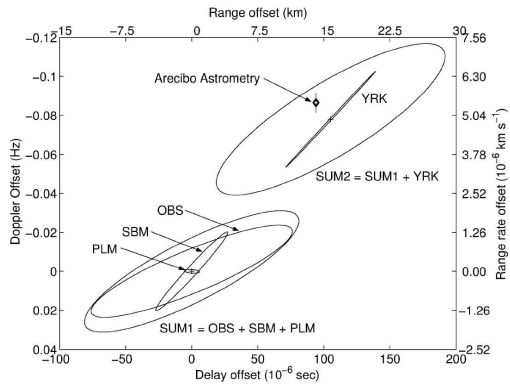}
\caption{ 
   Prediction error ellipses for the location in time delay (distance) 
   and Doppler shift (line-of-sight velocity) of the 0.5 km NEA 6489 
   Golevka for an Arecibo observation in 2003 based on not including 
   (SUM1) and including (SUM2) the non-gravitational force known as the 
   Yarkovsky effect. The actual measurement indicated by "Arecibo
   astrometry" clearly shows that the Yarkovsky effect is important in
   modifying the orbits of small bodies \citep[from][]{ches03}.
   }
\label{yarkovsky}
\end{figure}

Astrometry and characterization would be major objectives of an SKA
based radar system. NEAs are of great interest due to their potential
hazard to the Earth, as objectives for future manned space missions to
utilize their resources and as clues to the early history of the solar
system.  Astrometry and measurements of their sizes and spin vectors
will greatly reduce the uncertainties in projecting their future orbits
including non-gravitational influences such as the Yarkovsky effect
(Figure~\ref{yarkovsky}).  Measurements of the shapes, sizes and
densities will provide insights as into the internal structure of NEOs,
important both for understanding their history and also for designing
mitigation methods should an NEO pose a significant threat to Earth.
Unambiguous surface imagery at resolutions of a few meters will give
insights into their collisional histories while the polarization
properties of the reflected echo can be used to detect the presence of
regoliths.  Shapes, sizes and surface structure are currently obtained
from multiple aspect angle delay-Doppler images
\citep[Figure~\ref{golevka} and][]{huds93}.  A radar equipped SKA will
have the capability to image NEOs out to about 0.3 AU from Earth
allowing large numbers to be imaged at resolutions of less than 20 m.
The current Arecibo 13 cm radar system has the capability to image NEOs
with about 20 m resolution to distances of approximately 0.05 AU. With
over 100 times Arecibo's current sensitivity, an SKA based radar system
could achieve similar resolutions at 0.15 to 0.20 AU and much higher
resolutions for closer objects. The synthesized beam of the SKA
(assuming 3,000 km baseline and 10 GHz frequency) has a spatial
resolution at 0.2 AU of about 300 m, very much larger than the
achievable resolution based on the sensitivity but small enough to
mitigate the effects of delay-Doppler ambiguities allowing improved
shape modeling and surface imagery.  Doppler discrimination in the
synthesis imagery will provide the plane-of-sky direction of the
rotation vector \citep{depa94} and polarization properties will
elucidate regolith properties. 

\begin{figure}[tbh]
\centering
\includegraphics[width=0.2\textwidth]{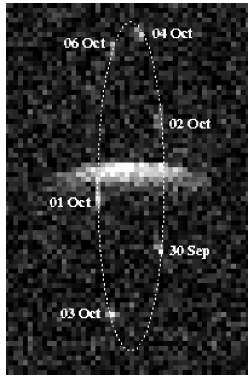}
\caption{ A composite Arecibo delay-Doppler image of the binary near
   Earth asteroid 2000 DP107 showing the primary body with the location 
   of the secondary on the dates shown in 2000.  The diameters of the 
   two bodies are about 800 m and 300 m, the orbital radius and period 
   are 2.6 km and 1.76 days, respectively, giving a density for the 
   primary of approximately 1.7 g cm$^{-3}$ \citep{marg02b}. Figure 
   courtesy of J.L. Margot.
   }
\label{binNEA}
\end{figure}

Because of their implications for both the composition and internal
structure of asteroids, measurements of densities would be a major
objective of SKA observations of NEAs. The discovery of binary NEAs
\citep[Figure~\ref{binNEA};][]{marg02b} provided the first opportunity
for direct measurements of densities for the 10-20\% of NEAs that are
estimated to be in binary configurations. However, while they provide
important information about NEA densities, the primary and secondary
components of these binaries are a particular class of NEAs
\citep{marg02b} and are not fully representative of the general
population.  An alternative method of estimating densities for NEAs is
the measurement of the Yarkovsky effect via long term astrometric
observations \citep{vokr04}.  The size of the effect is dependent on
the spin rate, the thermal inertia of the surface and the mass. The
first two of these can be measured or estimated allowing the mass to be
estimated and, hence, the density if the asteroidÕs volume is known via
a shape model. 

\subsubsection{Main Belt Asteroids}

\begin{figure}[tbh]
\centering
\includegraphics[width=0.45\textwidth]{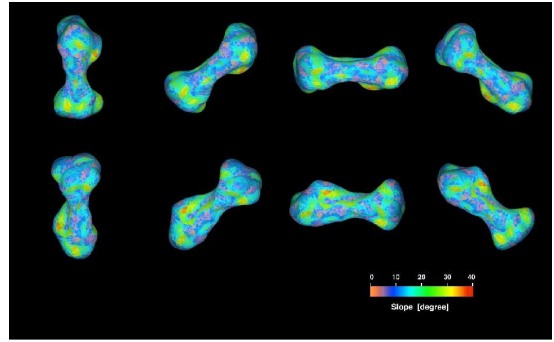}
\caption{ Shape models for the metallic main belt asteroid 216
   Kleopatra derived from Arecibo delay-Doppler radar images. The model
   shows Kleopatra to be 217 km x 94 km x 81 km. It may be the remains 
   of a collision of two former pieces of an ancient asteroids 
   disrupted core.  The color coding indicates the gravitational slopes 
   \citep{ostr00}.
   }
\label{kleopatra}
\end{figure}

A SKA based radar system would have a unique ability to measure the
properties of small bodies out to the far edge of the main asteroid
belt; sizes, shapes, albedoes and orbital parameters. The current
Arecibo radar system has only been able to obtain a shape model for one
MBA, Kleopatra \citep[Figure~\ref{kleopatra};][]{ostr00} and measure
the radio wavelength reflection properties of a relatively small number
of asteroids near the inner edge of the belt \citep{magr01} plus those
for a few of the very largest MBAs such as Ceres and Vesta (M. Nolan
private communication). Main belt issues that a SKA based radar could
address are: 1) The size distribution of MBAs would provide valuable
constraints on material strength and, hence, on collisional evolution
models; 2) Measurement of proportion of MBAs that are in binary systems
would provide information about the collisional evolution of the main
belt and detection of these systems would also provide masses and
densities for a large number of MBAs; 3) Astrometry would also provide
masses and densities via measurements of the gravitational perturbation
from nearby passes of two bodies and also, for small bodies, from
measurements of the Yarkovsky effect; 4) From radar albedo measurements
determine whether there is a switch within the main belt from rocky to
icy objects and, if so, whether it is gradual or abrupt. 

\subsubsection{Comets}

Spacecraft flybys have provided reasonable detailed information about
three comets, Halley, Borelly and Wild, and over the next 1-2 decades, 
prior to the completion of the SKA, a small number of additional comets 
will be studied from spacecraft such as the already launched Deep 
Impact and Rosetta missions and from potential new missions such as a 
successor to the failed Contour mission. Direct measurements of the 
sizes of three comets have indicated that cometary nuclei have very low 
optical albedoes and this has led to an upward revision of the size 
estimates of comets based on measurements of their absolute magnitudes.
However, the very small sample means that the distribution of cometary
albedoes is very uncertain and, hence, there is still considerable
uncertainty as to the size distribution of comets. An SKA based radar
system could resolve this issue which has ramifications related to the
assumed source of short period comets in the Kuiper belt.
\citet{bern04} have pointed out that there is a significant shortage of
KBOs at small sizes if comets have nuclei that are in the 10 km range
as currently thought. A SKA based radar would be able to image cometary
nuclei out to about 1 AU obtaining sizes, shapes, rotation vectors, and
actual nucleus surface images. For objects at larger distances, size 
estimates will be obtained from range dispersion and also from rotation 
periods from radar light curves combined with measurements of Doppler 
broadening. Over time these measurements would be the major source of 
cometary size estimates.

\subsection{Technical issues}

For many bodies, unambiguous plane-of-sky synthesis imagery is superior
to delay-Doppler imagery. Consequently, for both imaging and
astrometric observations, a SKA based radar would need to have the
capability to do both traditional radar delay-Doppler observations and
synthesis imaging of radar illuminated objects. For both range-Doppler
imaging and astrometric observations of near earth objects, range
resolutions of 20 ns or better will be required. At 10 GHz the angular
resolution of even the proposed central compact array will be smaller
than the angular size of some NEOs requiring that for delay-Doppler and
astrometric observations the SKA will still need to be used in an
imaging mode. While adding to the complexity of the observations, the
small spatial extent of the synthesized beam will greatly assist in
mitigating the ambiguities inherent in delay-Doppler imaging.
Delay-Doppler processing will require access to the complex outputs of
the correlator at a 20 ns or better sampling rate. This requirement may
not be dissimilar from that required for pulsar observations but it
will have the added complexity that the NEOs will be in the near field
of the SKA.

\section{Extrasolar Giant Planets}

The detection of extrasolar giant planets is one of the most exciting 
discoveries of astronomy in the past decade.  Despite the power of the 
radial velocity technique used to find these planets, it is biased 
to finding planets which are near their primary and orbiting edge-on.  
To augment those planets found by radial velocity searches, detections 
using astrometry, which are most sensitive to planets orbiting face-on, 
are needed.  Many researchers are eagerly searching for ways to 
directly detect these planets, so they can be properly characterized 
(only the orbit and a lower limit to the mass is known for most 
extrasolar planets). Below we will discuss potential contributions for 
SKA. 

\subsection{Indirect detection by astrometry}

The orbit of any planet around its central star causes that star to 
undergo a reflexive circular motion around the star-planet barycenter.  
By taking advantage of the incredibly high resolution of SKA, we may be 
able to detect this motion.  Making the usual approximation that the 
planet mass is small compared to the stellar mass, the stellar orbit 
projected on the sky is an ellipse with angular semi-major axis 
$\theta_r$ (in arcsec) given by:
\begin{equation}
   \theta_r = {m_p \over M_*} \, {a_{AU} \over D_{pc}} \quad , 
\end{equation}
where $m_p$ is the mass of the planet, $M_*$ is the mass of the star, 
$a_{AU}$ is the orbital distance of the planet (in AU), and $D_{pc}$ is 
the distance to the system (in parsecs).

The astrometric resolution of SKA, or the angular scale over which 
changes can be discriminated ($\Phi$), is proportional to the intrinsic 
resolution of SKA, and inversely proportional to the signal to noise 
with which the stellar flux density is detected (${\rm SNR}_*$):
\begin{equation}
   \Phi = { \theta_{HPBW} \over {2 \cdot {\rm SNR}_*}} \quad . 
\end{equation}
This relationship provides the key to high precision astrometry: the 
astrometric accuracy increases both as the intrinsic resolution 
improves and also as the signal to noise ratio is increased.  
Astrometry at radio wavelengths routinely achieves absolute astrometric 
resolutions 100 times finer than the intrinsic resolution, and can 
achieve up to 1000 times the intrinsic resolution with special care.  
As long as the phase stability specifications for SKA will allow such 
astrometric accuracy to be achieved for wide angle astrometry, such 
accuracies can be reached.

When the astrometric resolution is less than the reflexive orbital 
motion, that is, when $\Phi \ \lsim \ \theta_r $, SKA will detect 
that motion.  We use the approximation that $\theta_{HPBW} 
\sim \lambda / B_{max}$, so that detection will occur when:
\begin{equation}
   {\rm SNR}_* \ \gsim \ 10^5 \, {\lambda \over B_{max}} 
\left( {m_p \over M_*} \, {a_{AU} \over D_{pc}} \right)^{-1} \quad .
\end{equation}
The factor of $2 \times 10^5$ enters in to convert from radians 
to arcseconds.  

Note, however, that astrometric detection of a planet requires that 
curvature in the apparent stellar motion be measured, since linear 
terms in the reflex motion are indistinguishable from ordinary stellar 
proper motion.  This implies that at the very minimum, one needs three 
observations spaced in time over roughly half of the orbital period of 
the observed system.   A detection of a planetary system with 
astrometry would thus require some type of periodic monitoring.  

We use the technique described in \citet{butl03a} to calculate the 
expected flux density from stars, and whether we can detect their 
wobble from the presence of giant planets.  If all of the detectable 
stars for SKA (roughly 4300), had planetary companions, how many of 
them could be detected (via astrometry) with SKA?

We assume that the planets are in orbits with semimajor axis of 5 AU.
We consider 3 masses of planetary companions: 5 times jovian, jovian, 
and neptunian.  We assume integration times of 5 minutes, at 22 GHz.  
From the Hipparcos catalog \citep{perr97}, there are $\sim$ 1000 stars 
around which a 5*jovian companion could be detected, $\sim$ 620 stars 
around which a jovian companion could be detected, and $\sim$ 40 stars 
around which a neptunian companion could be detected.  Virtually 
none of these stars are solar-type.  From the Gliese catalog 
\citep{glie88}, there are $\sim$ 1430 stars around which a 5*jovian 
companion could be detected, $\sim$ 400 stars around which a jovian 
companion could be detected, and $\sim$ 60 stars around which a 
neptunian companion could be detected.  Of these, $\sim$ 130 are 
solar analogs.

\subsection{Direct detection of gyro-cyclotron emission}

Detection of the thermal and synchrotron emission from Jupiter, taken 
to distances of the stars, is beyond even the sensitivity of SKA unless 
prohibitively large amounts of integration time are spent.  However, 
Jupiter experiences extremely energetic bursts at long wavelengths.  If 
extrasolar giant planets exhibit the same bursting behavior, SKA might 
be used to detect this emission.  If such a detection occurred, it 
would provide information on the rotation period, strength of the 
magnetic field, an estimate of the plasma density in the magnetosphere, 
and possibly the existence of satellites.  The presence of a magnetic 
field is also potentially interesting for astrobiology, since such a 
field could shield the planet from the harsh stellar environment.  Some 
experiments have already been done to try to detect this emission 
\citep{bast00}.

These bursts come from keV electrons in the magnetosphere of the 
planet.  The solar wind deposits these electrons, which can 
subsequently develop an anisotropy in their energy distribution, 
becoming unstable.  When deposited in the auroral zones of the planet, 
emission results at the gyrofrequency of the magnetic field at the 
location of the electron ($f_g = 2.8 B_{gauss}$ MHz, for magnetic 
field strength $B_{gauss}$ in G).  This kind of emission occurs on 
Earth, Saturn, Jupiter, Uranus, and Neptune in our solar system.  The 
emission can be initiated or modulated by the presence of a satellite 
(Io, in the case of Jupiter).  

If we took the mean flux density of Jupiter at 30 MHz ($\sim$ 50000 Jy 
at 4.5 AU) to 10 pc, the resultant emission would only be 0.2 $\mu$Jy.  
This is very difficult to detect, given the expected sensitivity of 
SKA at the lowest frequencies.  However, the emission is variable (over 
two orders of magnitude), some EGPs may have intrinsically more 
radiated power, and if the emission is beamed, there is a significant 
increase in the expected flux density. 

The details of the expected radiated power from this emission mechanism 
are outlined in \citet{farr99} and \citet{zark01}.  We summarize the 
discussion here.  There exists a very good correlation amongst those 
planets that emit long wavelength radio waves between radiated power 
and input kinetic power from the solar wind.  Given expressions for the 
solar wind input power and conversion factor, and a prediction of the 
magnetic moment of a giant planet, we can write the expected radiated 
power as:
\begin{equation}
   P_{rad} \sim 400 \left(\omega \over \omega_j\right)^{4 \over 5} 
                \left(M \over M_j\right)^{4 \over 3} 
                \left(d \over d_j\right)^{8 \over 5} 
                \ [{\rm GW}] \ ,
\end{equation}
where $\omega$, $M$, and $d$ are the rotational rate, mass, and 
distance to primary of the planet, and the subscripted $j$ quantities 
are those values for Jupiter.  The expected received flux density can 
then be easily calculated, assuming isotropic radiation.

The frequency at which the power is emitted is limited at the high 
end by the maximum gyrofrequency of the plasma: $f_g \sim 2.8 
B_{gauss}$ [MHz], for magnetic field strength $B_{gauss}$ in G.  This 
usually limits such emissions to the 10's of MHz (Jupiter's cutoff is 
$\sim$40 MHz), but in some cases (for the larger EGPs), can extend into 
the 100's of MHz.  For this reason, these kinds of experiments might be 
better done with LOFAR, but there is still a possibility of seeing some 
of them at the lower end of the SKA frequencies.

If we take the current list of EGPs and use the above formalism to 
calculate the expected flux density, we can determine which are the 
best candidates to try to observe gyrocyclotron emission from.  In 
this exercise, we exclude those planets with cutoff frequencies $<$ 10 
MHz (Earth's ionospheric cutoff frequency), and those in the galactic 
plane (because of confusion and higher background temperature).  
Table~\ref{egp} shows the top four candidates, from which it can be 
seen that the maximum predicted emission is of order a few mJy (note 
that \citet{farr99} found similar values despite using slightly 
different scaling laws).  But, again, this is the mean emission, so 
bursts would be much stronger, and beaming could improve the situation 
dramatically.  Given the multi-beaming capability of SKA, it would be 
productive to attempt monitoring of some of the best candidates for 
these kinds of outbursts in an attempt to catch one.

\begin{table}[tbh]
\caption{Four best candidates for EGP gyrocyclotron emission detection. }
\label{egp}
\vspace*{1truemm}
\footnotesize
\begin{center}
\begin{tabular}{ccc}
\hline\hline
\noalign{\vspace{3pt}}
Star & $f_g$ (MHz) & $F_r$ (mJy) \\
\noalign{\vspace{3pt}}
\hline
\noalign{\vspace{3pt}} 
$\tau$ Bootes & 42  & 4.8 \\
Gliese 86   & 44  & 2.3 \\
HD 114762   & 202 & 0.28 \\
70 Vir      & 94  & 0.13 \\
\noalign{\vspace{3pt}}
\hline\hline
\end{tabular}
\end{center}
\end{table}

\section{The Sun}

The Sun is a challenging object for aperture synthesis, especially
over a wide frequency range, due to its very wide range of spatial
scales (of order 1 degree down to 1\arcsec), its lack of fine
spatial structure below about 1\arcsec, its great brightness
(quiet Sun flux density can be $10^6$--$10^7$ Jy) and variability
(flux density may change by 4-5 orders of magnitude in seconds),
and its variety of relevant emission mechanisms (at least
three---bremsstrahlung, gyroemission, and plasma emission---occur
regularly, and others may occur during bursts).  The key to
physical interpretation of solar radio emission is the analysis of
the brightness temperature spectrum, and because of the solar
variability this spectrum must be obtained over relatively short
times (less than 1 s for bursts, and of order 10 min for slowly
varying quiescent emission).  This means that broad parts of the
RF spectrum must be observed simultaneously, or else rapid
frequency switching must be possible.  The Sun produces only
circularly polarized emission--any linear component is destroyed
due to extreme Faraday rotation during passage through the corona.
High precision and sensitivity in circular polarization
measurements will be extremely useful in diagnostics of the
magnetic field strength and direction.

Through long experience with the VLA and other instruments, it has
been found that only antenna spacings less than about 6 km are
useful, which corresponds to a synthesized beam of
$10\arcsec/\nu_{\rm GHz}$. This empirical finding agrees with
expectations for scattering in the solar atmosphere \citep{bast94}.

Given the specifications of the SKA, some unique solar science can
be addressed in niche areas, but only if the system takes account
of the demands placed on the instrument as mentioned above. For
flares, the system should be designed with an ALC/AGC time
constant significantly less than 1 s, should allow for rapid
insertion of attenuation, and should allow for rapid frequency
switching. There will be little use for the beam-forming (phased
array) mode, since even very low sidelobes washing over the Sun
will dominate the signal, and there is no way to predict where a
small beam should be placed to catch a flare.  In synthesis mode,
the main advantage of SKA will be in its high sensitivity to low
surface brightness variations.  The following solar science could
be addressed:

\subsection{Solar bursts and activity}

The Frequency Agile Solar Radiotelescope (FASR) will be designed
to do the best possible flare-related science, and it is hard to
identify unique science to be addressed by SKA in this area.
However, if SKA is placed at a significantly different longitude
than FASR, it can cover the Sun at other times and produce useful
results. To cover the full Sun, small antennas (of order 2 m above
3 GHz, and 6 m below 3 GHz) are required.  Larger antenna sizes,
while restricting the field of view, can also be useful when
pointed at the most flare-likely active region.

\subsection{Quiet sun magnetic fields}

The magnetic geometry of the low solar atmosphere governs the
coupling between the chromosphere/photosphere and the corona.
Hence, it plays an important role in coronal heating, solar
activity, and the basic structure of the solar atmosphere. One can
uniquely measure the magnetic field through bremsstrahlung
emission of the chromosphere and corona, which is circularly
polarized due to the temperature gradient in the solar atmosphere.
At $\nu\lsim10$ GHz, bremsstrahlung is often swamped by
gyroemission, but it dominates at higher frequencies over much of
the Sun.  By measuring the percent polarization $P\%$ and the
local brightness temperature spectral slope $n = -\partial \log
T_b/\partial\log\nu$, one can deduce the longitudinal magnetic
field $B_\ell = (107/n\lambda)P\%$, with $B_\ell$ in G \citep{gelf04}.  
To reach a useful range of field strengths, say 10 G, the polarization 
must be measured to a precision of about 0.1--0.2\% (since $n$ is 
typically between 1 and 2).

Both FASR and EVLA will address this science area, but FASR's
small (2 m) antennas mean that the complex solar surface will have
to be imaged over the entire disk with high polarization
precision, while EVLA's relatively small number of baselines will
make imaging at the required precision difficult.  If SKA has
relatively large antennas (20 m) and high polarization precision,
it will be able to add significantly to this important
measurement.

\subsection{Coronal Mass Ejections}

Coronal Mass Ejections (CMEs) are an important type of solar
activity that dominates conditions in the interplanetary median
and the Sun's influence on the Earth.  Understanding CME
initiation and development in the low solar atmosphere is critical
to efforts to understand and predict the occurrence of CMEs. It is
expected that CMEs can be imaged through their bremsstrahlung
emission, but such emission will be of low contrast with the
background solar emission.  \citet{bast97} determined that
the best contrast should occur near 1 GHz.  Although one of the
FASR goals is to observe CMEs, the nearly filled aperture and high
sensitivity of SKA to low contrast surface brightness variations
can make it very sensitive to CMEs.  In addition to following the
temporal development of the CME morphology, SKA spectral
diagnostics can constrain the temperature, density, and perhaps
magnetic field within the CME and surrounding structures.

\section*{Acknowledgements}

Comments from Jean-Luc Margot, Mike Nolan, and Steve Ostro were 
appreciated.

\end{document}